\documentclass[12pt,twoside, epsf]{article}

\usepackage{color,graphicx,times}

\makeatletter
\long\def\M#1{\leavevmode\setbox\@tempboxa\hbox{#1}\@tempdima\fboxrule
    \advance\@tempdima \fboxsep \advance\@tempdima \dp\@tempboxa
   \hbox{\lower \@tempdima\hbox
  {\vbox{\hrule \@height \fboxrule
          \hbox{  \hskip\fboxsep
          \vbox{\vskip\fboxsep \box\@tempboxa\vskip\fboxsep}\hskip
                 \fboxsep\vrule \@width \fboxrule}%
                  }}}}

\makeatother

\let \ttorg \tt \def \tt{\ttorg \obeyspaces}

\begin{document}

\title{\Large\bf Self-Reference, Biologic and the Structure of Reproduction }
\author{Louis H. Kauffman\\ Department of Mathematics, Statistics \\ and Computer Science (m/c
249)    \\ 851 South Morgan Street   \\ University of Illinois at Chicago\\
Chicago, Illinois 60607-7045\\ $<$kauffman@uic.edu$>$\\}

\maketitle

\thispagestyle{empty}

\subsection*{\centering Abstract}

{\em In this paper we explore the boundary shared by biology and formal systems.}
\bigbreak

\noindent {\bf Keywords.} logic, algebra, topology, biology, replication, celluar automaton, quantum, DNA, container, extainer


\section{Introduction} 

This paper concentrates on relationships of formal systems with biology. In particular, this is a study of different forms
and formalisms for replication. The paper is based on previous papers by the author \cite{KL,BL1,BL2}. We have freely used texts of those papers where the formulations are of use, and we 
have extended the concepts and discussions herein considerably beyond the earlier work. We concentrate here on formal systems not only for the sake of showing how there is a fundamental mathematical structure to biology, but also to consider and reconsider philosophical and phenomenological points of view in relation to natural science and mathematics. The relationship with phenomenology
\cite{Hei,MP,H,Hide,Badiou,Rosen} comes about in the questions that arise about the nature of the observer in relation to the observed that arise in philosophy, but also in science in the very act of determining the context and models upon which it shall be based. Our original point of departure was cybernetic epistemology \cite{UU,FV,GSB,MUV,REQ,REF,EF,EF1,VL,SS,SRF,BL1,BL2,KL}
and it turns out that cybernetic epistemology has much to say about the relation of the self to structures that may harbor a self. It has much to say about the interlacement of selves and organisms.
This paper can be regarded as an initial exploration of this theme of mathematics, formalities, selves and organisms - presented primarily from the point of view of cybernetic epistemology, but with the intent that these points of view should be of interest to phenomenologists. We hope to generate fruitful interdisciplinary discussion in this way.
\\

Our point of view is structural. It is not intended to be reductionistic. There is a distinct difference between building up structures in terms of principles and imagining that models of the world are constructed from some sort of building-bricks. The author wishes to make this point as early as possible because in mathamatics one naturally generates hierarchies, but that does not make the mathematician a reductionist. We think of geometry as the consequences of certain axioms for the purpose of organizing our knowledge, not to insist that these axioms are in any way other than logically prior to the theorems of the system. Just so, we look for fundamental patterns from which certain complexes of phenomena and ideas can be organized. This does not entail any assumption about ``the world'' or how the world may be built from parts. Such assumptions are, for this author, useful only as partial forms of explanation.\\

We examine the schema behind the reproduction of DNA. As all observers of science know, the pattern of the DNA reproduction is very simple.
The DNA molecule consists of two interwound strands, the Watson Strand (W)  and the Crick Strand (C). The two strands are bonded to each other via a backbone of base-pairings and these bonds can be broken by certain enzymes present in the cell. In reproduction of DNA the bonds between the two strands are broken and the two strands then acquire the needed complementary base molecules from the cellular environment to reconstitute each a separate copy of the DNA. At this level the situation can be described by a symbolism like this.
$$DNA = <W|C> \longrightarrow <W|\,\, E\,\, |C> \longrightarrow <W|C> \, <W|C> = DNA \,\, DNA.$$ Here $E$ stands for the environment of the cell. The first arrow denotes the separation of the DNA into the two strands. The second arrow denotes the action between the bare strands and the environment that leads to the production of the two DNA molecules.\\

Much is left out of this schema, not the least of which is the ignoring of the word {\it interwound}. Indeed the DNA molecule is a tight spiral winding of its two interlocked strands and so the new DNA's would be linked around one another if it were not for the work of other enzymes that mysteriously manage to unlink the new DNA's in time for cell division to occur. We discuss this briefly in Section 2 of the present paper. Nevertheless, this is the large scale description of the replication of DNA that is fundamental to the division of cells and to the continuance of living organisms.\\

The abstract structure of this DNA replication schema makes it a pivot to other models and other patterns. To see this most clearly, suppose we have $O$ and $O^{*}$ algebraic entities such that 
$O^{*} O = 1$ where $1$ denotes an algebraic identity such that $1A = A1 = A$ for any other algebraic entity $A.$ Assume that  juxataposition (multiplication) of algebra elements is associative. Let $P = OO^{*}.$ Then $$P =  OO^{*} =  O 1 O^{*} =  OO^{*}OO^{*} = PP.$$ Thus we have a general algebraic form for the self-replication described above. Note that in algebra we do not choose a direction of
change. Thus we have $1 = O^{*}O.$ In the replication scenario this is replaced by a {\it process arrow} $$1 \longrightarrow O^{*}O$$ generalizing the arrow $$E \longrightarrow |C> \, <W|$$ where the environment $E$ can supply Crick and Watson strands (via the base pairing) to the opened DNA. Thus algebra provides a condensed formalism for discussing self-replication. See Section 9.1 of the present paper for examples (via the Temperley-Lieb algebra) that follow these algebraic patterns.
\\

In the DNA formalism above, we can reverse the roles of $C$ and $W$ and use instead of $DNA = <W|C> = <W|\,|C>$, $$DNA = |C><W|$$ and the dual assumption that the environment $E$ is like an identity
element in context where $E = <W|C>.$ Then we would have 
$$DNA = |C> <W| \longrightarrow  |C>\,\, E\,\, <W|\longrightarrow |C> \, <W|C> \,<W|$$
$$ = |C> \, <W|\,|C> \,<W| =DNA \,\, DNA.$$
We can choose either pattern as is convenient. The reader will find that we use both of these formalisms in the paper.\\

We now invite the reader to examine the form of the science involved in this well-known description. We speak of the DNA molecules as though we could see them directly in the phenomenology of our ordinary sight. Some science does involve the direct extension of sight as the experience of looking through a telescope or a light microscope. But in the case of the DNA one proceeds by logical consistency and the indirect but vivid images via the electron microscope and the patterns of gel electrophoresis. In the case of electron microscope images there is every reason to assume (that is, it appears consistent to assume) that the objects shown can be taken to be analogous to the macroscopic objects of our perception. This means that one has the possibility of observing ``directly" that DNA molecules can be knotted. I do not say that one can observe directly the coiling of the Watson and the Crick strands, but the full DNA can be observed as though it were a long rope. This rope can be seen to be coiled and knotted in electron micrographs such as the one shown here in Figure~\ref{micrograph}. Even this ``showing'' requires a difficult technique beyond the usual techniques of the electron microscope. The DNA was coated with protein by the experimenters so that it became a chain of larger and more robust diameter. Then the electron microscope revealed the patterns of knotting in an apparent projection of the coated DNA from three dimensional space to the two dimensional space of the image.\\

 Scientists strive to make this information consistent and repeatable. This means that whether or not a scientist believes that the microworld of the DNA is just like our world of objects, he can nevertheless assent to the facts shown by the observations that {\ if} we assume object behavior similar to our realm for the DNA realm, then these instruments reveal knotting and other forms of geometric patterning.
 A phenomenologist can criticize the lack of direct perception in this form of science, but in fact it is remarkable how consistent is the hypothesis of indirect perception on which the work is based.
 Most working biologists would not question the basis of their biological perceptions direct or indirect. But they would instantly question the bases of the experiments and their consistency.
 For those who are philosophically inclined there is a lesson to be learned here about experimental phenomenology \cite{Hide}. One wants to know how far a world-view can be extended before it disintegrates. A phenomenological theme is illustrated here in that what we see in the electron micrograph is deeply shaped by the complex story of biological experiment that surrounds it.
This is a deep example of the same type shown in more elementary circumstances by Hide \cite{Hide} in discussion the Necker Cube and multiple interpretations associated with it and dependent on stories related to it. In the body of the paper, I shall make other phenomenological remarks about the various aspects and themes of the paper. \\
 
 Along with these forays into experimentation, there are also analogous forays into the limits of logic. Here we meet the replication schema again. Replication in logic is intimately related to self-reference and to formalisms that, if not properly interpreted, can lead to paradox. The reasons for this are, by now, apparent. The usual mathematical formalisms for set theory assume that there is no temporal evolution in the structures. The sets do not change over time. A set like the Russell {\it set of all sets that are not members of themselves} crosses the boundary of such restrictions. Once the Russell set is declared, the set itself comes under scrutiny for the very property that defines it. In this case, if we think recursively, the new Russell set is not a member of itself, but it is a new set, just created. And so we must take a step, and form a new Russell set that includes the first one. This new Russell set is also subject to scrutiny and must be further included in a yet again new Russell set.
 The process continues ad infinitum. A declaration of set membership has led to a recursive process of self-production. This may look like a tragedy for the classical mathematics, but it is exactly what interests us when studying biology!  Mathematical Biology is concerned with those abstract structures leading to recursive generation of structures from themselves and from their environments. For this reason we explore such abstract schema in this paper.\\
 
 Looked at from the structure of process, the biological phenomenology takes a different shape from the one we sketched in terms of experiment and observation. Now we are engaged in a mathematical phenomenology, looking for those places in mathematics where the temporal and the atemporal coexist in a balance that can be perturbed in favor of process but can also be held in favor of stability.
 It is such structures that can stand in back of organisms that survive in the world. One can imagine a future biology that derives the classification of viable organisms from such a mathematical theory
 of morphogenesis. Rene Thom \cite{Thom} may have had just such an idea behind his work ``Structural Stability and Morphogenesis." Exploration of Thom's work in relation to our ideas will be the subject of another paper.\\

A simplest form of recursive replication is formalized by the following consideration: Suppose that we have a domain $D$ 
(a {\it reflexive domain}) where every element $a \in D$ is also seen as a mapping from $D$ to $D,$ $$a:D \longrightarrow D.$$
We will write $ab$ for the application of the mapping $a$ to an element $b$ of $D.$ It is assumed that mappings of $D$ to itself that are expressed by algebraic formulae in the terms of this composition $ab$ are themselves represented by elements of $D.$ Thus given a formula such as $((ax)b)(xc)$ there exists an element $F \in D$ such that $Fx = ((ax)b)(xc)$ for all $x \in D.$
This seemingly innocuous property has a train of consequences. For example, suppose that $a$ is any element of $D.$ Define a new element $G \in D$ by the equation
$$Gx = a(xx).$$ (Our reflexive assumption guarantees the existence of such a $G.$) Then $$GG = a(GG).$$ One can read this equation as saying that 
{\it every $a \in D$ has a fixed point of the form $GG$ for an element $G \in D.$} The element $GG$ is {\it productive} in that it produces an $a$ and will continue to do so.
$$GG \longrightarrow a(GG) \longrightarrow a(a(GG)) \longrightarrow a(a(a(GG))) \longrightarrow \cdots.$$ 
Here the arrow is intended to indicate the process of production of the $a$. $GG$ is like a cell that can divide, and once it divides it can divide again. It is not lost upon us that $GG$ as an abstract cell
is its own genetic material and {\it doubled} (two $G$'s) in an abstract hint of the double helix of $DNA.$\\

The mathematical phenomenology of this fixed point construction can be illustrated by a shift of notation. Define $$Gx = \langle xx \rangle.$$ As long as $x$ is not $G,$ then this operator seems quite 
innocuous, but when we allow $x=G$ then we have $$GG = \langle GG \rangle,$$ and the form $GG$ has miraculously appeared inside itself. The notational shift is effective when the reader takes on the brackets as an enclosure, for then he can be surprised that a form would enclose itself. I intend to give the reader a phenomenological shock of this kind by using the shift of notation. The shift is not necessary for the shock, but anyone who sees the Church-Curry fixed point construction and is not shocked, has not seen the story to its roots. Here is the problem of understanding laid bare. We can tell a joke, but will the listener get the joke? What does it mean to get the joke?\\

I say that $GG$ is an {\it eigenform} for the operator
$T(x) = \langle x \rangle$ since it is a fixed point for that operator. See \cite{REQ,REF,EF,EF1,VL,SS,SRF} and note that a fixed point $V$ such that $T(V) = V$ is analogous to an eigenvector with eigenvalue equal to one. Thus one can think of the 
eigenforms associated with a given transformation as correpondent to a generalized {\it spectrum} of the operator $T.$
Eigenforms go beyond numerical spectra to Fixed Points in larger domains.
It is still spectral analysis of a kind, but Eigenforms speak to the arising of `objects as tokens for eigenbehaviours' where an eigenbehaviour is a behaviour that has 
the character of a fixed point even when that fixed point is in a newly created domain not part of the status quo of the transformations that engendered it.
There are many examples. For example, Heinz von Forerster points out that the sentence ``I am the observed relation between myself and observing myself.'' defines the concept of ``I'' as an eigenform of the transformation\\

 $T(x) =$ ``The observed relation between $x$ and observing $x.$'' \cite{UU}. \\
 
\noindent  In the arising of a solution to the equation 
$$I = T(I),$$  an ``I''  comes into being.
These `I' are not part of the status quo of the systems that engender them. 
They are transcendent to those systems, and are often seen as illusory or otherwise magical. 
But one can also regard the `I' as the direct result of the action of the organism itself. Thus here we find a nexus that allows many points of view, from the classical transcendental view of the self to the 
intertwined phenomenological view of self and world in mutual embrace and mutual creation as in the work of Merleau-Ponty \cite{MP}.
Observing systems can have ``I''s but they do not produce them. They are them.
This says a great deal about the efficacy of using cybernetic epistemology to understand understanding. We hope that the reader will bear with these attempts at comparison that will surely become more systematic in later work. These comparisons are important particularly in facing the question of how organisms acquire awareness and how awareness can be applied to itself.\\

\noindent {\bf Remark.} In a purely formal treatment or in a computer program, one must take care of the possibility of uncontrolled recursion. It is worth noting that at the linguistic level, there is no intent to repeat in ``I am the one who says I.".  Another example of this ``stopping" is the famous sentence due to Quine: \\

\noindent {\it Refers to itself when appended by its own quotation ``refers to itself when appended to its own quotation."} \\

\noindent There is no necessity for an uncontrolled recursion to occur at the point of self-reference or self-replication. It is a matter of context. In the case of DNA reproduction the replication happens in the cell only under very special conditions, and it is immediately followed by the separation of the new DNA's into their respective new cells. The new cells can then undergo mitosis again, but that self-replication is dependent upon the possibilities in the environment.\\

\begin{figure}
     \begin{center}
     \begin{tabular}{c}
     \includegraphics[width=6cm]{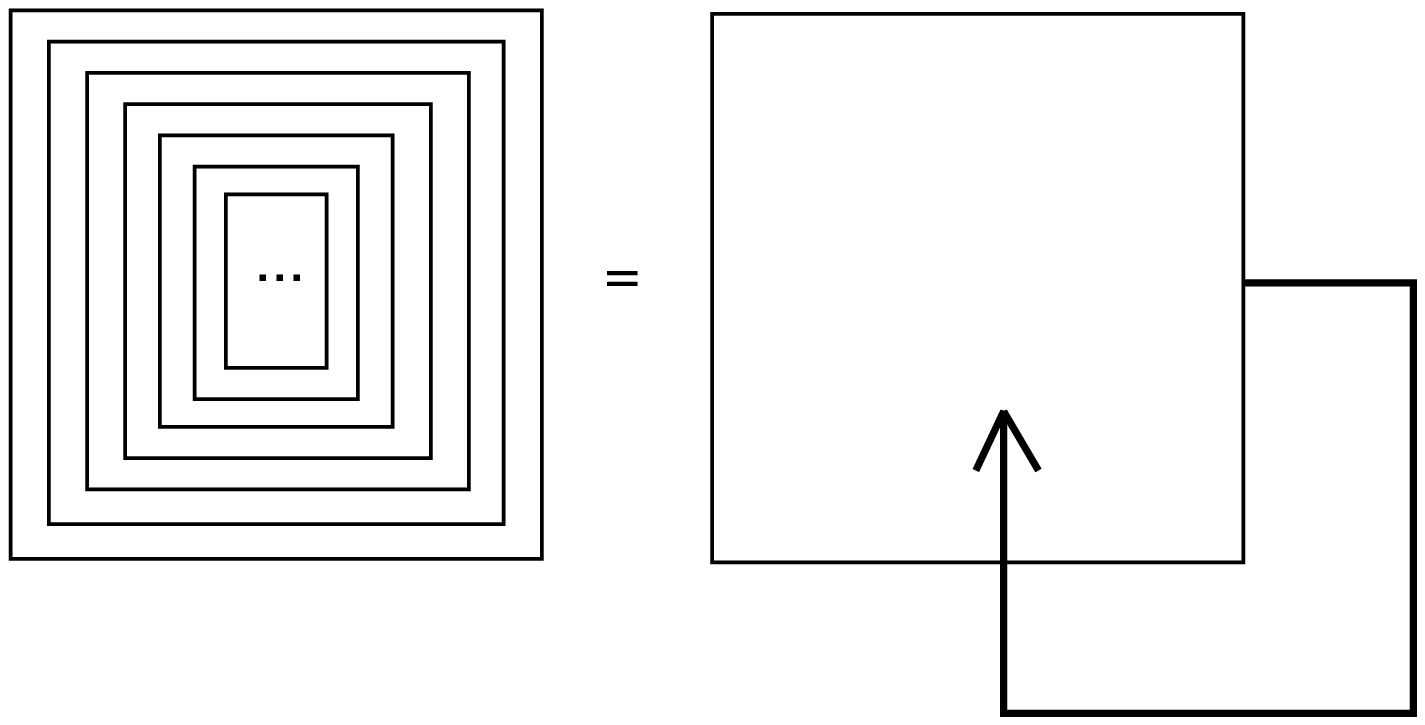}
     \end{tabular}
     \caption{\bf An Infinite Fixed Point}
     \label{fix}
\end{center}
\end{figure}

 Theoretically, uncontrolled recursion leads to the notion of fixed points in a direct manner by talking the limit of iterated recursion. 
 Consider the transformation $$F(X) = \M{X}.$$
 If we iterate it  and take the limit we find 
 $$G = F(F(F(F( \cdots )))) = \M{\M{\M{\M{ ... }}}}$$ 
 an infinite nest of marks satisfying the equation
 $$G = \M{G}.$$
 With $G = F(G),$ I say that $G$ is an {\em eigenform} for the transformation $F,$ and this is an eigenform that occurs by taking a limit of the recursion.
 See Figure~\ref{fix} for an illustration of this nesting with boxes and an arrow that points inside the reentering mark to indicate its appearance inside itself. If one thinks of the mark itself as a Boolean logical value, then extending the language to include the reentering mark $G$ goes beyond the boolean. We will not detail here how this extension can be related to non-standard logics, but refer the reader to \cite{KL}.  \bigbreak
 
The interesting thing about such recursive processes is that in the abstract they do not stop. The abstraction that never stops is behind the actual processes that stop eventually due to lack of resources or other means. We will consider mathematical patterns of replication in Section 4 of the present paper. Just as one may consider such abstractions, one can idealize process by taking it to infinity. In the case of $GG$ we see that we can consider a formal fixed point of $a$ in the form 
$$P = a(a(a(a(a(a(a(a(a(a(a(a(a(a(a(\cdots))))))))))))))).$$ In a microcosm, such patterns are the fractal residue of recursive processes in organisms. They do not reach ideal infinity in actuality, but the ideal 
lies behind the real and has its own mathematical reality. The reader can see how, in the cybernetic epistemology, the statement that the ideal lies behind the real takes the form of the interlacement 
of recursion and the generation of the object as a token for the behavior of that recursion. Without an interaction of subject and object neither subject nor object can come forth into a world, nor can that world come forth. This is not to say that there is no background to these circularities. It is a question of attending to process and mutuality to realize the creative nature of the world into which the subject is thrown and the nature of the subject into which the world is thrown.\\

Interpretation of this basic semantic/syntactic level yields untold riches. For example, consider the von Neumann schema for a Universal Building Machine denoted $B.$ $B$ will produce any entity $X$ for which there is a blueprint $x.$ We write $$B,x \longrightarrow X, x.$$ The machine $B$ produces the entity $X$ with a copy of its blueprint $x$ attached. Now let $b$ be the blueprint for $B$ itself.
Then we have $$B,b \longrightarrow B,b.$$ The machine $B$ has reproduced itself. The {\it pattern} of this reproduction is the same as the pattern for the reflexive fixed point. We might have written 
$$Bx \longrightarrow xx$$ and $$BB \longrightarrow BB.$$ The separation of entity and blueprint is our distinction. In biology the two strands of DNA are each blueprint and entity. Nature begins without the logical distinctions that we find so compelling.\\

Reflexive domains are daring in their nature. They dare to allow process and form to coexist. They dare to combine time and timelessness. To see the fine wire we are wallking, the reader should consider a reflexive domain that allows the operation of negation $\sim$. Such a domain could be one that allows logical discourse among its operators. We then define 
$$Rx = \sim xx ,$$ and find that $$RR = \sim RR .$$ Thus we find an element $RR$ in this logical reflexive domain that is its own negation. Such objects are not allowed in classical logical domains.
To see this circularity to its bitter end, interpret $Ax$ as ``$x$ is a member of $A$.". Then $R$ is identified as the ``set of all $x$ that are not members of themselves", and we see that we have reproduced the Russell Paradox. For us the way out is via the recursion. But this requires further discussion for which biology and its lessons can help.\\ 

In living systems there is an essential circularity that is
the living structure. Living systems produce themselves from themselves and the materials and energy of the environment.
There is a strong contrast in how we avoid circularity in mathematics and how nature revels in biological circularity.
One meeting point of biology and 
mathematics is knot theory and topology. This is no accident, since topology is indeed a controlled study of cycles and circularities
in primarily geometrical systems.  
\bigbreak

In this paper we will discuss DNA replication, logic and biology, the relationship of symbol and object, and the emergence of form. It is in 
the replication of DNA that the polarity (yes/no, on/off, true/false) of logic and the continuity of topology meet. Here 
polarities are literally fleshed out into the forms of life. The reader may wish to compare our thought  with the eloquent work and intuition of Steven Rosen \cite{Rosen} that also would add flesh to topology and find the phenomenological root of life and physicality.
\bigbreak

We shall pay attention 
to the different contexts for logic, from the mathematical to the biological to the quantum logical. In each case
there is a shift in the role of certain key 
concepts. In particular, we follow the notion of {\it copying} through these contexts and with it gain new insight into the role of replication
in biology, in formal systems and in the quantum level (where one cannot copy a state without erasing it!). 
\bigbreak

In the end we arrive at a summary formalism, a chapter in {\em boundary mathematics} (mathematics using directly
the concept and notation of containers
and delimiters of forms - compare \cite{WB} and \cite{GSB}) where there are not only containers $<>$, but also
extainers  $>< \, \, ,$  entities open to interaction and distinguishing 
the space that they are not. In this formalism we find a key for the articulation of diverse relationships. The 
{\em boundary algebra of containers and extainers} is to biologic what boolean algebra is to classical logic. Let $C \, = \, <>$ and $E \, = \,  ><.$ Then 
$$EE  \, = \,   ><><  \, = \,  >C<$$ and $$CC  \, = \,  <><>  \, = \,  <E>.$$ Thus an extainer produces a container when it interacts with itself, and a container produces an extainer
when it interacts with itself. The formalism of containers and extainers can be compared with Heidegger's lifeworld of objects sustaining each other through mutual transpermeation
\cite{Hei}, a mutual interpenetration that gives rise to form.
\bigbreak

The formalism of containers and extainers is a chapter in the foundations of a symbolic language for shape and interaction. 
With it, we can express the {\em form} of DNA replication succinctly as follows: Let the DNA itself be represented as a container

$$\mbox{\rm DNA} \, = \, <>.$$
 
\noindent We regard the two brackets of the container as representatives for the two matched DNA strands. We let the extainer $E=><$ represent the 
cellular environment with its supply of available base pairs (here symbolized by the individual left and right brackets).  When the DNA
strands separate, they encounter the matching bases from the environment and become two DNA's.

$$ \mbox{\rm DNA} = \, <> \, \longrightarrow \, < E > \, \longrightarrow \, < > < > \, = \, \mbox{\rm DNA} \,\, \mbox{\rm DNA}.$$ 

\noindent
Life itself is about systems that search and learn and become. The little symbol  $$E \, = \, ><$$ with the property that 
$$EE \, = \, ><><$$ producing containers $<>$ and retaining its own integrity in conjunction with the autonomy of $<>$ (the DNA)
can be a step toward bringing formalism to life.
\bigbreak

The paper is organized in eleven sections. The first section is this Introduction. The second section  discusses DNA replication. The third  section discusses the logic structure of replication in general and 
DNA replication in particular. The fourth section discusses reflexive domains and lambda calculus. The fifth section discusses copying and the curious circumstance that (in the usual scientfic theory)  life is supported at the molecular level by structures just emerging from the quantum mechanical restriction on copies (the no-cloning theorem). Section 6 is a discussion of the mathematics of Laws of Form. Here we take the form of distinction for the form. Percept and concept are cradled in the form of distinction. Mathematics and Logic exfoliate from the form of distinction. Forms of replication and self-reference arise naturally as distinctions are both name and action in the form. Section 7 is a discussion of Knot Logic, models that are based on topological forms. Here we initiate a question about the fundamental nature of topology in all of our modeling, mathematical, physical and biological. Section 8 is an epistemological digression. It is a discussion between two characters, Cookie and Parable who are sentient text strings. This section allows us to range over and synthesize the many ideas that have been initiated in the paper so far. Section 9 discusses the mathematical and epistemological and topological basis for the constructions in this paper, and introduces extainers and containers as we have discussed them above. Section 10 discusses recursive distinguishing,  autopoesis and cellular automata. In Section 10 we show how the scheme of recursive distinguishing, discovered by Joel Isaacson \cite{JI} contains a self-replicating form that can be interpreted as either the abstract self-replication of containers and extainers, or as the division and integration of protocells in an elementary form of artificial life. Section 11 is an epilogue on the themes of the paper.\\

\section{Replication of DNA}
We start this essay with the question: During the replication of DNA, how do the daughter
DNA duplexes avoid entanglement? In the words of John Hearst \cite{KH}, we are in search of the mechanism for
the ``immaculate segregation".  This question is inevitably involved with the topology of the DNA, for
the strands of the DNA are interwound with one full turn for every ten base pairs. With the strands
so interlinked it would seem impossible for the daughter strands to separate from their parents.
\bigbreak

\begin{figure}
     \begin{center}
     \begin{tabular}{c}
     \includegraphics[width=6cm]{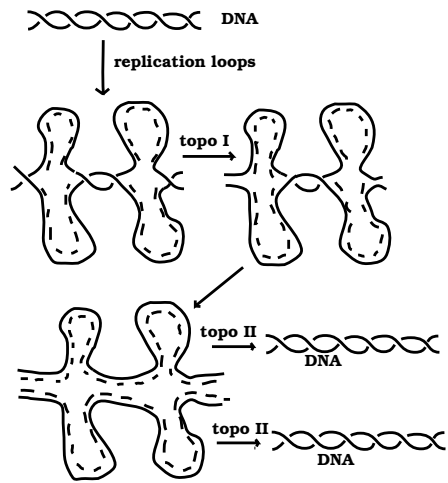}
     \end{tabular}
     \caption{\bf DNA Replication}
     \label{Figure 1}
\end{center}
\end{figure}

A key to this problem certainly lies in the existence of the topoisomerase enzymes that can change the
linking number between the DNA strands and also can change the linking number between two DNA duplexes.
It is however, a difficult matter at best to find in a tangled skein of rope the just right crossing 
changes that will unknot or unlink it.  The topoisomerase enzymes do just this, changing crossings by 
grabbing a strand, breaking it, and rejoining it after the other strand has slipped through
the break. Random strand switching is an unlikely mechanism, and one is led to posit some intrinsic
geometry that can promote the process. In \cite{KH} there is made a specific suggestion about this intrinsic 
geometry. It is suggested that {\em in vivo} the DNA polymerase enzyme that promotes replication (by creating 
loops of single stranded DNA by opening the double stranded DNA) has sufficient rigidity not to allow the new 
loops to swivel and become entangled. In other words, it is posited that the replication 
loops remain simple in their topology so that the topoisomerase can act to promote the formation of 
the replication loops, and these loops once formed do not hinder the separation of the newly born 
duplexes. The model has been to some degree confirmed \cite{LD, Pflumm}.  
In the first stages of the formation of the replication loops Topo\,I acts favorably to allow their 
formation and amalgamation. Topo\,II has a much smaller job of finishing the separation of the newly 
formed duplexes. In Figure~\ref{Figure 1} we illustrate the schema of this process. In this Figure we indicate the action of the Topo\,I
by showing a strand being switched in between two replication loops. The action of Topo\, II is only stated but not shown.
In that action, newly created but entangled DNA strands would be disentangled. Our hypothesis is that this second action is
essentially minimized by the rigidity of the ends of the replication loops {\em in vivo} and the fact that newly created DNA is quickly 
compacted in the cell, preventing further catenation (linking).
\bigbreak

In the course of this research, we started thinking about the diagrammatic logic of DNA replication
and more generally about the relationship between DNA replication, logic and basic issues in the 
foundations of mathematics and modeling. The purpose of this paper is to explain some of these 
issues, raise questions and place these questions in the most general context that we can muster at
this time. This paper is foundational.  It will not, in its present form,
affect issues in practical biology, but we hope that it will enable us and the reader to ask fruitful
questions and perhaps bring the art of modeling in mathematics and biology forward.
\bigbreak

There are many more questions. One takes as given the double helical structure of the DNA and the topological difficulties that ensue from this structure.
One takes as given the remarkable topological enzymes that Nature has provided to solve the problem. One wants to know more details about how the topological problem is solved. 
But beyond that, one would like to know if all this was necessary. Did Nature have to work with a helical structure that produced linked molecules without further aids? Could there have been another way that avoid the topological pitfalls? Why did the ideal of simple reproduction meet the difficulties of topology at this juncture? We are far from answers to these questions.\\

To this end we have called the subject matter of this paper ``biologic" with the intent that this 
might suggest a quest for the logic of biological systems or a quest for a ``biological logic" or 
even the question of the relationship between what we call ``logic" and our own biology. We 
have been trained to think of physics as the foundation of biology, but it is possible to 
realize that indeed biology can also be regarded as a foundation for thought, language, mathematics 
and even physics. In order to bring this statement over to physics one has to learn to admit that
physical measurements are performed by biological organisms either directly or indirectly and that
it is through our biological structure that we come to know the world. This foundational view will be 
elaborated as we proceed in this paper.\\

On the other hand, we are, in thinking about logical process and biology, continually faced with the fact that there is a ground level of processes in physics and biology that we cannot erase by calling them just  our descriptions. We meet actuality in the attempt to describe and understand an observed world. The same phenomenon occurs in mathematics. Once certain structures are in place ( for example, the natural numbers) then a host of facts and relationships ensue that are to all concerned, regarded as inevitable (for example that the product of two odd numbers is odd, or that there are infinitely many prime numbers). There is a bedrock to science and we are faced with biological structures that appear to lead inevitably to our own cognition. When we include ourselves in the biology, then we are saying that biological structures can think and, in particular, they can in full thought examine themselves. If we make this inclusion, then we have to admit that we do not have any way yet to construct a complete story from the nuts and bolts of the logic of DNA reproduction to the production of thought and thoughtfulness of the scientist who asks these questions. The essential circularity of the study is always present and it is important to understand and accept the circularity without creating unfounded systems of belief.\\

\subsection {DNA Recombination}
\begin{figure}
     \begin{center}
     \begin{tabular}{c}
     \includegraphics[width=6cm]{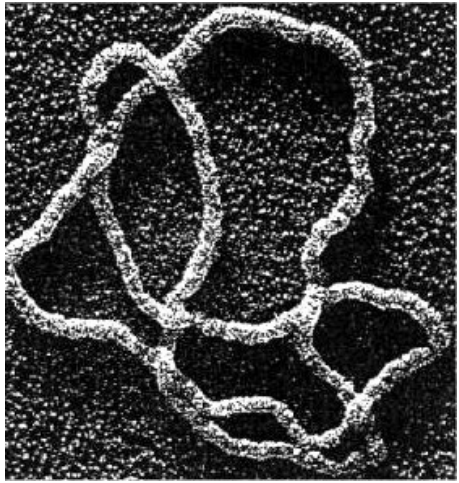}
     \end{tabular}
     \caption{\bf DNA Knot}
     \label{micrograph}
\end{center}
\end{figure}

In Figure~\ref{micrograph} we show an electron micrograph of a closed circular DNA molecule that has been coated with protein \cite{Cozz} in such a way that the knotting of the DNA is apparent from the micrograph. In fact, the electron micrograph has been made in such a way that a biologist can read it as a knot diagram.  The concept and application of  knot diagrams was essential for this scientific application.  Electron micrographs are two dimensional renderings. It was necessary that topological information in the DNA knot could be determined from such a projection . The knot diagram becomes a connection between the invisible world of the DNA molecules and the structural topological world of the mathematics. \\

\begin{figure}
     \begin{center}
     \begin{tabular}{c}
     \includegraphics[width=8cm]{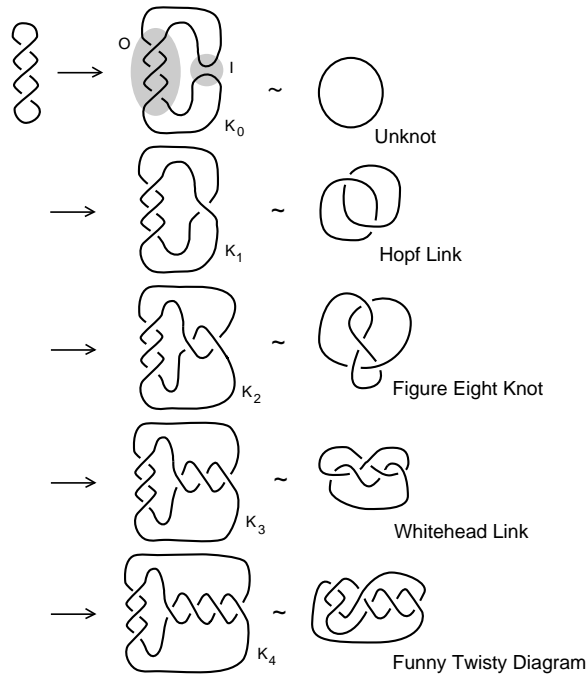}
     \end{tabular}
     \caption{\bf Recombination Process}
     \label{recomb}
\end{center}
\end{figure}

Now examine Figure~\ref{recomb}. In this figure we begin with an idealized bit of closed circular DNA, not knotted, but harboring a three-fold twist. Just after the arrow, the DNA is bent over so that two of its arcs are in proximity,  and its knot type is indicated as a 
closed circle to the right. At the second arrow a recombination event occurs.
The DNA is broken and re-spliced, forming a crossing of right-handed type. The result of the recombination is a simple link of two components with linking number equal to 1. A second recombination occurs and a figure-eight knot appears. Then comes a so-called Whitehead link and then a more complex knot. We see in this diagrammatic experiment the results of successive recombination, where the pattern of recombination (insertion of one right-handed twist) is always the same.
The diagrammatic experiment is linked (no pun intended) with real experiments and appropriate electron micrographs \cite{Cozz}  to show that a real processive recombination actually does produce these knots and links. The result is that the hypothesis of the form of the recombination is confirmed by this diagrammatic experiment  coupled with the surrounding knot theory and the surrounding molecular biology.  It is the language of the diagrams that provides the crucial connection between the biology and the mathematics.\\

In attempting to understand the processes that occur in the replication and combination of DNA we are working at the limits of our perception of objects that have the usual locatable properties of things that occur in our macroscopic world. It is assumed by the molecular biologist that the long chain molecule of the DNA can be approximated in its behavior by a long twisted rope or a chain of smaller objects interconnected in ways that are familiar. These long-chain molecules are composed of atoms whose bonding properties are best explained at the quantum mechanical level. Thus in studying DNA  we are examining the very locus of the emergence of the world of the familiar topological properties of knotted rope. This is confirmed by the DNA experiments in replication where the hypothesis that DNA behaves like knotted rope is confirmed in the predictions of the diagrammatic experiments in processive recombination.\\

Is there another way to look at this phenomenology? One might be skeptical of the topological properties of the DNA were it not for the vivid amplification of the electron micrographs and the consistency of the simple models of recombination with the results of experiment. Invisible worlds are hypothesized to behave in simple topological and geometrical ways. \\

\section{Logic, Copies, and DNA Replication}
In logic it is implicit at the syntactical level that copies of signs are 
freely available. In abstract logic there is no issue about materials available for the production of copies of
a sign, nor is there necessarily a formalization of how a sign is to be copied. In the practical realm there
are limitations to resources. A mathematician may need to replenish his supply of paper. A computer
has a limitation on its memory store. In biology, there are no signs, but there are entities that 
we take as signs in our description of the workings of the biological information process. In this 
category the bases that line the backbone of the DNA are signs whose significance lies in their 
relative placement in the DNA. The DNA itself could be viewed as a text that one would like to copy.
If this were a simple formal system it would be taken for granted that copies of any given text can
be made. Therefore it is worthwhile making a comparison of the methods of copying or reproduction that
occur in logic and in biology.
\bigbreak

In logic there is a level beyond the simple copying of symbols that contains a non-trivial description
of self-replication.  The (von Neumann) schema is as follows: There is a universal building machine $B$ that 
can accept a text or description $x$ (the program) and build what the text describes. We let lowercase
$x$ denote the description and uppercase $X$ denote that which is described.  Thus $B$ with $x$ will build
$X.$  The building machine also produces an extra copy of the text $x.$ This is 
appended to the production $X$ as $X,x.$ Thus $B$, when supplied with a description $x,$ produces that which $x$
describes, with a copy of its description attached. Schematically we have the process shown below.
\bigbreak

$$B,x \longrightarrow B,x ; X,x$$

\noindent
Self-replication is an immediate consequence of this concept of a universal building machine.
Let $b$ denote the text or program for the universal building machine.  Apply $B$ to
its own description.

$$B,b \longrightarrow B,b ; B,b$$

\noindent
The universal building machine reproduces itself. Each copy is a universal 
building machine with its own description appended. Each copy will proceed to reproduce itself
in an unending tree of duplications. In practice this duplication will continue until all available
resources are used up, or until someone removes the programs or energy sources from the 
proliferating machines.
\bigbreak

It is not necessary to go all the way to a universal building machine to establish 
replication in a formal system or a cellular automaton (See the epilogue to this paper for examples.).
On the other hand, all these logical devices for replication
are based on the hardware/software or Object/Symbol distinction. It is worth looking at the abstract 
form of DNA replication.
\bigbreak

DNA consists in two strands of base-pairs wound helically around a phosphate backbone.  It is 
customary to call one of these strands the ``Watson" strand and the other the ``Crick" strand.
Abstractly we can write  $$DNA = <W|C>$$ \noindent to symbolize the binding of the two strands into the single
DNA duplex. Replication occurs via the separation of the two strands via polymerase enzyme.
This separation occurs locally and propagates. Local sectors of separation can amalgamate into
larger pieces of separation as well. Once the strands are separated, the environment of the cell 
can provide each with complementary bases to form the base pairs of new duplex DNA's. Each strand,
separated {\em in vivo}, finds its complement being built naturally in the environment. This picture ignores
the well-known topological difficulties present to the actual separation of the daughter strands.
\bigbreak

The base pairs are $AT$ (Adenine and Thymine) and $GC$ (Guanine and Cytosine). Thus if
 
$$<W| = <...TTAGAATAGGTACGCG... |$$ \noindent then $$|C> = |...AATCTTATCCATGCGC...>.$$ 

\noindent Symbolically we can oversimplify the whole process as 

$$<W| + E \longrightarrow <W|C> = DNA$$

$$E + |C> \longrightarrow <W|C> = DNA$$

$$<W|C> \longrightarrow <W| + E + |C> = <W|C> <W|C>$$

\noindent Either half of the DNA can, with the help of the environment, become a full DNA.
We can let $E \, \longrightarrow \, |C><W|$ be a symbol for the process by which the environment supplies the 
complementary base pairs $AG$, $TC$ to the Watson and Crick strands. In this oversimplification
we have cartooned the environment as though it contained an already-waiting strand $|C>$ to 
pair with $<W|$ and an already-waiting strand $<W|$ to pair with $|C>.$ 
\smallbreak 

{\em In fact it is the opened
strands themselves that command the appearance of their mates. They conjure up their mates from 
the chemical soup of the environment.} 
\smallbreak 

The environment $E$ is an identity element in this algebra of 
cellular interaction. That is, $E$ is always in the background and can be allowed to appear spontaneously
in the cleft between Watson and Crick:

$$<W|C> \longrightarrow  <W| |C> \longrightarrow <W| E |C>$$ 

$$\longrightarrow <W| |C><W| |C> \longrightarrow <W|C><W|C>$$

\noindent This is the formalism of DNA replication.  
\bigbreak

Compare this method of replication with the movements of the universal building machine supplied 
with its own blueprint. Here Watson and Crick ( $<W|$ and $|C>$ ) are each both the machine {\em and} 
the blueprint for the DNA. They are complementary blueprints, each containing the information to
reconstitute the whole molecule. They are each machines in the context of the cellular 
environment, enabling the production of the DNA.  This coincidence of machine and blueprint, hardware
and software is an important difference between 
classical logical systems and the logical forms that arise in biology.
\bigbreak

\section{Lambda Algebra and Reflexive Domains}
One can look at formal systems involving self-replication that do not make a 
distinction between Symbol and Object. In the case of formal systems this means that one is 
working entirely on the symbolic side, quite a different matter from the biology where there is no
intrinsic symbolism, only our external descriptions of processes in such terms. An example 
at the symbolic level is provided by the lambda calculus of Church and Curry \cite{BD} where functions
are allowed to take themselves as arguments. This is accomplished by the following axiom. 
\bigbreak

\noindent {\bf Reflexive Axiom for Lambda Algebra}:  Let $A$ be an algebraic system with one binary operation denoted 
$ab$ for elements $a$ and $b$ of $A.$ Let $F(x)$ be an algebraic expression over $A$ with one variable $x.$
Then there exists an element $a$ of $A$ such that $F(x) = ax$ for all $x$ in $A.$ 
\bigbreak

In broad terms, the axiom states that
if a computation can be described using terms from the algebra $A,$ then there
exists an element of $A$ which performs that computation.
\bigbreak

\noindent {\bf Remark.} In the Introduction to this paper, we have formulated the lambda algebra in terms of a reflexive domain $D$ where elements of the domain are in one-to-one correspondence with algebraic mappings of the domain to itself. The reader should compare the two formulations and see that they are formally identical. The notion of reflexive domain is of serious importance for
the epistemology of extended logic, biology and cybernetics. Once we consider domains that include their own observers as in biological systems, social systems and indeed science itself where we admit that the scientist is part of his own study, then the territory under investigation becomes a reflexive domain. Here I use the term reflexive domain in a wider sense than just the aspect of algebraically defined mappings. The concept in the larger sense is dependent on the recognition that the actions of the observers in the space or territory affects and is part of what is being studied.
Mathematics is an excellent example of such a reflexive domain where the constructs and definitions of the mathematicians become new mathematics to be studied and extended by them. Note that in this case the actors (the mathematicians) are playing into their own study and making it in the very act of performing that study. Similarly biological organisms in the form of their actions produce the very biology that they are. This can all be said with varying degrees of emphasis upon an observer external to the process. We can take the stance that no observer is fully external to the process, but that some are construed to be more independent than others.\\

The reader will be interested in the paper \cite{BuligaKauff} in which we formulate a diagrammatic lambda caclulus that can be generalized to 
abstract chemical interactions (chemlambda in the paper and see other references within). The key here is local interaction in the sense of interactions of graphical nodes.
The relationships of chemlambda with the present paper are multifold, but require another paper to unfold. \\

An algebra (not associative) that satisfies the reflexive axiom is a representation of the
lambda calculus of Church and Curry.  Let $b$ be an element of $A$ and define $F(x) = b(xx).$  
By the axiom we have $a$ in $A$ such that  $ax = b(xx)$ for any $x$ in $A.$ In particular (and this is where
the ``function" becomes its own argument)  $$aa = b(aa).$$ Thus we have shown that for any $b$ in $A$, there exists
an element $x$ in $A$ such that $x = bx.$ Every element of $A$ has a ``fixed point."

\noindent
This conclusion has two effects. It provides a fixed point for the function $G(x) = bx$ and it creates the
beginning of a recursion in the form 

$$aa = b(aa) = b(b(aa)) = b(b(b(aa))) = ...$$

\noindent The way we arrived at the fixed point $aa$ was formally the same as the
mechanism of the universal building machine.  Consider that machine:
\bigbreak

$$B,x \longrightarrow X,x$$   

\noindent We have left out the repetition of the machine itself. You could look at this as a machine that 
uses itself up in the process of building $X.$ Applying $B$ to its own description $b$ we have
the self-replication

$$B,b \longrightarrow B,b.$$

\noindent The repetition of $x$ in the form $X,x$ on the right hand side of this definition of the 
builder property is comParabel with  

$$ax = b(xx)$$

\noindent with its crucial repetition as well. In the fixed point theorem, the arrow is replaced by an 
equals sign! Repetition is the core of self-replication in classical logic.
{\em This use of repetition assumes the possibility of a copy at the syntactic level, in order to produce
a copy at the symbolic level.} There is, in this pivot on syntax, a deep 
relationship with other fundamental issues in logic. In particular this same form of repetition
is in back of the Cantor diagonal argument showing that the set of subsets of a set 
has greater cardinality than the original set, and it is in back of the G$\ddot{o}$del Theorem on the 
incompleteness of sufficiently rich formal systems. The pattern is also in back of the production of 
paradoxes such as the Russell paradox of the set of all sets that are not members of themselves.
\bigbreak

There is not space here to go into all these relationships, but the Russell paradox will give 
a hint of the structure.  Let ``Ax" be interpreted as ``x is a member of A". Then $Rx = \sim xx$ can be 
taken as the definition of a set $R$ such that $x$ is a member of $R$ exactly when it is {\em not} the case that 
$x$ is a member of $x.$ Note the repetition of $x$ in the definition $Rx = \sim xx.$ Substituting $R$ for $x$
we obtain  $RR = \sim RR$, which says that $R$ is a member of $R$ exactly when it is not the case that 
$R$ is a member of $R.$ This is the Russell paradox. From the point of view of the lambda calculus, we have found a 
fixed point for negation.
\bigbreak

Where is the repetition in the DNA self-replication?  The repetition and the replication are no longer separated. The repetition 
occurs not syntactically, but directly at the point of replication. Note the device of 
pairing or mirror imaging.  $A$ calls up the appearance of $T$ and $G$ calls up the appearance of $C.$
$<W|$ calls up the appearance of $|C>$ and $|C>$ calls up the appearance of $<W|.$  Each object $O$ calls up 
the appearance of its {\em dual or paired object} $O^*$.  $O$ calls up $O^*$ and $O^*$ calls up $O.$  The object that 
replicates is implicitly a repetition in the form of a pairing of object and dual object.
\smallbreak

\noindent $OO^*$ replicates via

$$O \longrightarrow OO^*$$

$$O^* \longrightarrow OO^*$$

\noindent whence

$$OO^* \longrightarrow O \,\,\, O^* \longrightarrow OO^* \,\,\, OO^*.$$

\noindent The repetition is inherent in the replicand in the sense that the dual of a form is a repetition of 
that form. The reproduction of  $OO^*$ is crucially dependent upon the possibility that $O$ and $O^*$ can separate to prepare for the condition that 
each will produce its complement. Thus we must distinguish between $OO^*$ and $O \,\,\, O^*$ and this distinction must come to pass. This is the first hint of the structure 
of a union and interaction of opposites that will be articulated further in a sequel to this paper.\\

\section{Quantum Mechanics, Copies and Distinctions}
We now consider the quantum level. It is common to speak as though the world were constructed from `lower' levels in a hierarchical process. In speaking this way, molecules are composed of atoms and atoms of more elementary particles. By the time we get to such elementary particles we are in the quantum level and a realm that does not any longer behave according to classical laws.
This is not the only way to speak about quantum mechanics. In fact our classical world is pervaded by quantum phenomena (the discrete spectral lines of hydrogen) and the distinction that needs to be made is when we observe such phenomena and how the notion of object must be changed in relation to our theories of such phenomena. One could declare that {\it there is no quantum world, but there are quantum theories that describe well certain classes of observation.} What we are about to describe is how the notion of an object is quite different in a quantum theory. In particular, for quantum states,
copying is not possible without the destruction of the original state.\\

We shall detail this matter of copying in a subsection. For a quantum process to copy a state, one needs a
unitary transformation to perform the job. One can show, as we will explain in section 5.2, that this 
cannot be done. There are indirect ways that seem to make a copy, involving
a classical communication channel coupled with quantum operators (so called quantum teleportation \cite{Lom}).
The production of such a quantum state constitutes a reproduction of the original
state, but in these cases the original state is lost, so teleportation looks more like transportation than copying.  
With this in mind it is fascinating to contemplate that DNA and other molecular configurations
are actually modeled (in principle) as certain complex quantum states.  At this stage we meet the
boundary between classical and quantum mechanics where conventional wisdom finds it is most useful to regard the main level of  
molecular biology as classical.
\bigbreak

We shall quickly
indicate the basic principles of quantum mechanics.  The quantum information context 
encapsulates a concise model of quantum theory:
\bigbreak

{\em The initial state of a quantum process is a vector $|v>$ in a complex vector space $H.$
Observation returns basis elements $\beta$ of $H$ with probability 

$$|<\beta \,|v>|^{2}/<v \,|v>$$

\noindent where $<v \,|w> = v^{*}w$ with $v^{*}$ the conjugate transpose of $v.$
A physical process occurs in steps $|v> \longrightarrow U\,|v> = |Uv>$ where $U$ is a unitary linear transformation.
\bigbreak

Note that since $<Uv \,|Uw> = <v \,|w>$ when $U$ is unitary, it follows that probability is preserved in the 
course of a quantum process.  }
\bigbreak

One of the details for any specific quantum problem is the nature of the unitary 
evolution.  This is specified by knowing appropriate information about the classical physics that 
supports the phenomena. This information is used to choose an appropriate Hamiltonian through which the 
unitary operator is constructed via a correspondence principle that replaces classical variables with appropriate quantum
operators. (In the path integral approach one needs a Langrangian to construct the action on which the path
integral is based.) One needs to know certain aspects of classical physics to 
solve any given quantum problem.  The classical world is known through our biology. In this sense 
biology is the foundation for physics.
\bigbreak

A key concept in the quantum information viewpoint is the notion of the superposition of states.
If a quantum system has two  distinct states $|v>$ and $|w>,$ then it has infinitely many states of the form
$a|v>+b|w>$ where $a$ and $b$ are complex numbers taken up to a common multiple. States are ``really" 
in the projective space associated with $H.$ There is only one superposition of a single state $|v>$ with 
itself. 
\bigbreak

Dirac \cite{D} introduced the ``bra-(c)-ket" notation $<A\,|B>= A^{*}B$ for the inner product of complex vectors $A,B \in H$.
He also separated the parts of the bracket into the {\em bra} $<A\,|$ and the {\em ket} $|B>.$ Thus

$$<A\,|B> = <A\,|\,\,|B>$$

\noindent In this interpretation,
the ket $|B>$ is identified with the vector $B \in H$, while the bra $<A\,|$ is regarded as the element dual to $A$ in the 
dual space $H^*$. The dual element to $A$ corresponds to the conjugate transpose $A^{*}$ of the vector $A$, and the inner product is 
expressed in conventional language by the matrix product $A^{*}B$ (which is a scalar since $B$ is a column vector). Having separated the bra and the ket, Dirac can write the
``ket-bra"  $|A><B\,| = AB^{*}.$ In conventional notation, the ket-bra is a matrix, not a scalar, and we have the following formula for the 
square of $P = |A><B\,|:$

$$P^{2} =  |A><B\,| |A><B\,| = A(B^{*}A)B^{*} = (B^{*}A)AB^{*} = <B\,|A>P.$$

\noindent Written entirely in Dirac notation we have
    
$$P^{2} =  |A><B\,| |A><B\,| =  |A><B\,|A><B\,|$$ 

$$= <B\,|A>\,|A\,><B| = <B\,|A>P.$$

\noindent The standard example is a ket-bra $P = |A\,><A|$ where $<A\,|A>=1$ so that $P^2 = P.$  Then $P$ is a projection matrix, 
projecting to the subspace of $H$ that is spanned by the vector $|A>$. In fact, for any vector $|B>$ we have 

$$P|B> = |A><A\,|\,|B> =  |A><A\,|B> = <A\,|B>|A>.$$

\noindent If $\{|C_{1}>, |C_{2}>, \cdots |C_{n}> \}$ is an orthonormal basis for $H$, and $P_{i} = |C_{i} \,><C_{i}|,$
then for any vector $|A>$ we have

$$|A> = <C_{1}\,|A>|C_{1}> + \cdots + <C_{n}\,|A>|C_{n}>.$$

\noindent Hence 

$$<B\,|A> = <C_{1}\,|A><B\,|C_{1}> + \cdots + <C_{n}\,|A><B\,|C_{n}>$$

$$ = <B\,|C_{1}><C_{1}\,|A> + \cdots + <B\,|C_{n}><C_{n}\,|A>$$

$$ = <B\,|\,\,\,[|C_{1}><C_{1}\,| + \cdots + |C_{n}><C_{n}\,|]\, \,\,|A>$$

$$ = <B\,|\,1^* \,\,|A>.$$

\noindent We have written this sequence of equalities from $<B\,|A>$ to $<B\,|1^* \,|A>$ to emphasize the role of the identity $1^*$ in the 
space of endomorphisms of the vector space $H:$

$$\Sigma_{k=1}^{n} P_{k} = \Sigma_{k=1}^{n} |C_{k}><C_{k}\,| = 1^*$$

\noindent so that one can write

$$<B\,|A> = <B\,|\,1^* \,|A> = <B\,| \Sigma_{k=1}^{n} |C_{k}><C_{k}\,| |A> = \Sigma_{k=1}^{n} <B\,|C_{k}><C_{k}\,|A>.$$

In the quantum context
one may wish to consider the probability of starting in state $|A>$ and ending in state $|B>.$ The 
square of the probability for this event is equal to $|<B\,|A>|^{2}$. This can be refined if we have more knowledge. 
If it is known that one can go from $A$ to $C_{i}$ ($i=1,\cdots,n$)
and from $C_{i}$ to $B$ and that the intermediate states $|C_{i}>$ are a complete set of orthonormal alternatives then we can assume that 
$<C_{i}\,|C_{i}> = 1$ for each $i$ and that $\Sigma_{i} |C_{i}><C_{i}| = 1^*.$  This identity now corresponds to the fact that
$1$ is the sum of the probabilities of an arbitrary state being projected into one of these intermediate states.
\bigbreak

If there are intermediate states between the intermediate states this formulation can be continued
until one is summing over all possible paths from $A$ to $B.$ This becomes the path integral expression 
for the amplitude $<B|A>.$
\bigbreak

\subsection{Quantum Formalism and DNA Replication}
We wish to draw attention to the remarkable fact that this formulation of the expansion of 
intermediate quantum states has exactly the same pattern as our formal summary of DNA replication.
Compare them. The form of DNA replication is shown below. Here the environment of possible base pairs is represented by the ket-bra
$E = |C><W \,|:$

$$<W|C>\, \longrightarrow \, <W|\, |C>\, \longrightarrow \,<W| E |C>$$ 

$$\longrightarrow\, <W|\, |C><W|\, |C>\, \longrightarrow\, <W|C><W|C>.$$

\noindent Here is the form of intermediate state expansion:

$$<B\,|A>\, \longrightarrow\, <B\,|\,|A>\, \longrightarrow \, <B\,|\,1 \,|A>$$

$$ \longrightarrow \, <B\,|\,\, \Sigma_{k}\,|C_{k}><C_{k}\,|\,\, |A> \, \longrightarrow \,\Sigma_{k}<B\,|C_{k}><C_{k}\,|A>.$$

\noindent We compare $$E = |C><W\,|$$ \noindent and $$1^* = \Sigma_{k}\,|C_{k}><C_{k}\,|.$$

\noindent That the unit $1^*$ can be written as a sum over the intermediate states is an expression of how the 
environment (in the sense of the space of possibilities) impinges on the quantum amplitude, just as the expression of the environment as a 
soup of bases ready to be paired (a classical space of possibilities) serves as a description of the biological environment.
The symbol $E = |C><W\,|$ indicated the availability of the bases from the environment to form the 
complementary pairs. The projection operators $|C_{i}><C_{i}\,|$ are the possibilities for interlock of 
initial and final state through an intermediate possibility. In the quantum mechanics the special pairing is not of bases but
of a state and a possible intermediate from a basis of states. It is through this common theme of 
pairing that the conceptual notation of the bras and kets lets us see a correspondence between such
separate domains.
\bigbreak

\subsection{Quantum Copies are not Possible}
Finally, we note that in quantum mechanics it is not possible to copy a quantum state!
This is called the no-cloning theorem of elementary quantum mechanics \cite{Lom}. Here is the proof: 
\smallbreak

\noindent{\bf Proof of the No Cloning Theorem.} In order to have a quantum process make a copy of a quantum
state we need a unitary mapping $U:H \otimes H \longrightarrow H \otimes H$ where $H$ is a complex vector space such that there is a fixed state
$|X>\, \in H$ with the property that $$U(|X>|A>) = |A>|A>$$ \noindent for any state $|A> \in H.$ (Note that $|A>|B>$ here denotes
the tensor product $|A> \otimes |B>.$) Let
$$T(|A>) = U(|X>|A>) = |A>|A>.$$ \noindent Note that $T$ is a linear function of $|A>.$  Thus we have 

$$T|0> = |0>|0> = |00>,$$

$$T|1> = |1>|1>=|11>,$$

$$ T(\alpha |0> + \beta |1>) = (\alpha |0> + \beta |1>)(\alpha |0> + \beta |1>).$$ 

\noindent But

$$ T(\alpha |0> + \beta |1>) = \alpha |00> + \beta |11>.$$  \noindent Hence

$$\alpha |00> + \beta |11> =(\alpha |0> + \beta |1>)(\alpha |0> + \beta |1>)$$

$$ = \alpha^{2} |00> + \beta^{2} |11> + \alpha \beta |01> + \beta \alpha |10>$$

\noindent From this it follows that 
$\alpha \beta = 0.$ Since $\alpha$ and $\beta$ are arbitrary complex numbers, this is a contradiction. $//$
\bigbreak

The proof of the no-cloning theorem depends crucially on the linear superposition of quantum states and the linearity of quantum process.
By the time we reach the molecular level and attain the possibility of copying DNA molecules we are copying in a quite different sense than the 
ideal quantum copy that does not exist. The DNA and its copy are each quantum states, but they are different quantum states! That we see the two 
DNA molecules as identical is a function of how we filter our observations of complex and entangled quantum states. Nevertheless, the identity
of two DNA copies is certainly at a deeper level than the identity of the two letters ``i" in the word identity. The latter is conventional and symbolic.
The former is a matter of physics and biochemistry. 
\bigbreak

Where does it happen that `things' become sufficiently separated that they can be copied? Where does it happen that distinctions cannot any longer be made in the firm sense of our idealized classical worlds? These are the philosophical and phenomenological questions that lie behind the differences between and the relations between classical quantum mechanics. We have raised these questions here. We have also shown how the Dirac formalism of bras and kets, so important to the quantum models, have a direct analogy with the self-replication of DNA.\\

\section{Laws of Form}
\begin{figure}
     \begin{center}
     \begin{tabular}{c}
     \includegraphics[height=4cm]{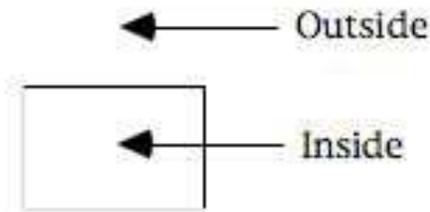}
     \end{tabular}
     \end{center}
     \caption{\bf Inside and Outside}
     \label{outin}
     \end{figure}

In this section we discuss a formalism due the G. Spencer-Brown \cite{GSB} that is often called the ``calculus of indications". This calculus is a study of mathematical foundations with a topological notation based on one symbol, the mark:
$$\M{ } \, .$$
This single symbol represents a distinction between its own inside and outside.
As is evident from Figure~\ref{outin}, the mark is regarded as a shorthand for a rectangle drawn in the plane and dividing the plane into the regions inside and outside the rectangle.   
The mark is seen as making a distinction and the calculus of indications is a calculus of distinctions where the mark refers to the act of distinction. Thus the mark is self-referential and refers to its own action and to the distinction that is made by the mark itself. Spencer-Brown is quite explicit about this identification of action and naming in the conception of the mark, and by the end of the book he reminds the reader that ``the mark and the observer are, in the form, identical". Thus the mark is an inherently phenomenological entity deriving its existence and its role from the participation of the reader or observer of the text that is the mark itself. That text is not an abstract text but an actual making of a distinction in the plane of the paper on which the mark appears, and in the mind of the reader/observer. We make this discussion here because it is important to trace the origins of the text such as the DNA strings that are fundamental to biology. Like the mark they are found by observation and come to be abstractions that guide our understanding of organism as a whole. Just so, and in greater generality, the mark becomes a pivot for the fundamental distinctions upon which we base any universe constructed by us in the course of our living.
\bigbreak

     \bigbreak

In the calculus of indications the mark can interact with itself in two possible ways. The resulting formalism becomes a version of Boolean arithmetic, but fundamentally simpler than the usual Boolean arithmetic of $0$ and $1$ with its 
two binary operations and one unary operation (negation).  \\

Is there  a {\it linguistic} particle that is its own anti-particle? Certainly we have
$$\sim \sim Q = Q$$ for any proposition $Q$ (in Boolean logic). And so we might 
write 
$$\sim \sim \longrightarrow *$$
where $*$ is a neutral linguistic particle, an identity operator so that
$$*Q = Q$$ for any proposition $Q.$ But in the normal use of negation there is no way that the negation sign combines with itself to produce itself.  Remarkably, the calculus of indications provides a context in which we can say exactly that a certain logical particle, the mark,  can act as negation {\it and} can interact with itself to produce itself.
\bigbreak

\begin{figure}
     \begin{center}
     \begin{tabular}{c}
     \includegraphics[height=4cm]{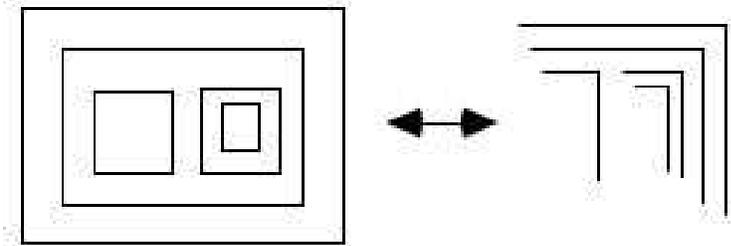}
     \end{tabular}
     \end{center}
     \caption{\bf Boxes and Marks}
     \label{boxmark}
     \end{figure}

In the calculus of indications patterns of non-intersecting marks (i.e. non-intersecting rectangles) are called {\it expressions.} For example in Figure~\ref{boxmark} we see how patterns of boxes correspond to patterns of marks.

     \bigbreak

\begin{figure}
     \begin{center}
     \begin{tabular}{c}
     \includegraphics[height=4cm]{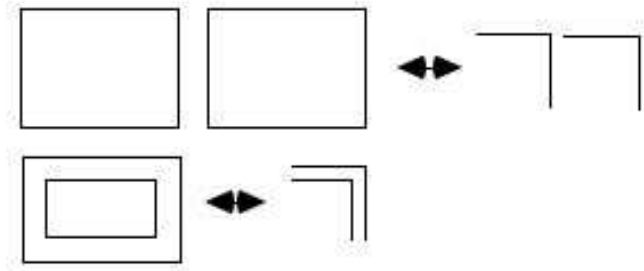}
     \end{tabular}
     \end{center}
     \caption{\bf Translation between Boxes and Marks}
     \label{markbox}
     \end{figure}

In Figure~\ref{boxmark}, we have illustrated both the rectangle and the marked version of the expression.  In an expression you can say definitively of any two marks whether one is or is not inside the other.  The relationship between two marks is either that one is inside the other, or that neither is inside the other.  These two conditions correspond to the two elementary expressions shown in
Figure~\ref{markbox}.

     \bigbreak

The mathematics in Laws of Form begins with two laws of transformation about these two basic expressions. Symbolically, these laws are:
\begin{enumerate}
\item Calling : $$\M{} \, \M{} \, =   \M{}$$  
\item Crossing: $$\M{ \M{ } }  =  \,\,\,\,.$$
\end{enumerate}
The equals sign denotes a replacement step that can be performed on instances of these patterns
(two empty marks that are adjacent or one mark surrounding an empty mark).
In the first of these equations two adjacent marks condense to a single mark, or a single mark expands to form two adjacent marks.  In the second equation  two marks, one inside the other, disappear to form the unmarked state indicated by nothing at all. 
That is, two nested marks can be replaced by an empty word in this formal system.  Alternatively, the unmarked state can be replaced by two nested marks. These equations give rise to a natural calculus, and the mathematics can begin.  For example,  {\it any expression can be reduced uniquely  to either the marked or the unmarked state.}  The he following example illustrates the method:
$$     \M{\M{\M{\M{} \M{}} \M{}} \M{}} \M{}  =   \M{\M{\M{\M{}} \M{}} \M{}}\M{} =   \M{\M{ \M{}} \M{}}\M{} $$
$$ = \M{\M{}}\M{} = \M{} \,\,\,.$$
The general method for reduction is to locate marks that are at the deepest places in the expression
(depth is defined by counting the number of inward crossings of boundaries needed to reach the given mark). Such a deepest mark must be empty and it is either surrounded by another mark, or it is adjacent to an empty mark. In either case a reduction can be performed by either calling or crossing. 
\bigbreak 

Laws of Form begins with the following statement.
``We take as given the idea of a distinction and the idea of an indication, and that it is not possible to make an indication without drawing a distinction. We take therefore the form of distinction for the form."  
Then the author makes the following two statements (laws):
\begin{enumerate}
\item {\it The value of a call made again is the value of the call.}
\item {\it The value of a crossing made again is not the value of the crossing.}
\end{enumerate}
The two symbolic equations above correspond to these statements. First examine the law of calling. It says that the value of a repeated name is the value of the name. In the equation
$$\M{} \, \M{} \, = \M{}$$
one can view either mark as the name of the state indicated by the outside of the other mark.  
In the other equation
$$\M{ \M{ } } = \,\,\,\,.$$
the state indicated by the outside of a mark is the state obtained by crossing from the state indicated on the inside of the mark. Since the marked state is indicated on the inside, the outside must indicate the unmarked state.  The Law of Crossing indicates how opposite forms can fit into one another and vanish into nothing, or how nothing can  produce opposite and distinct forms that fit one another, hand in glove.  The same interpretation yields the equation
$$\M{} \, = \, \M{}$$
where the left-hand side is seen as an instruction to cross from the unmarked state, and the right hand side is seen as an indicator of the marked state. The mark has a double carry of meaning. It can be seen as an operator, transforming the state on its inside to a different state on its outside, and it can be seen as the name of the marked state. That combination of meanings is compatible in this interpretation.  
\bigbreak

From the calculus of indications, one moves to algebra.  Thus 
 $$\M{\M{A}}$$
stands for the two possibilities
  $$\M{\M{\M{}}} \, = \, \M{} \,  \longleftrightarrow  \, A = \M{}$$
$$\M{\M{}} \, = \, \, \, \,  \longleftrightarrow \, A \,  = $$
In all cases we have
$$\M{\M{A}} \, = \, A.$$
 
 By the time we articulate the algebra, the mark can take the role of a unary operator
 $$ A \longrightarrow \M{A}.$$ But it retains its role as an element in the algebra.
Thus begins algebra with respect to this non-numerical arithmetic of forms.  The primary algebra that emerges is a subtle precursor to Boolean algebra.  One can translate back and forth between elementary logic and primary algebra:
\begin{enumerate}
\item $\M{} \longleftrightarrow T$
\item $\M{\M{}} \longleftrightarrow F$
\item $\M{A} \longleftrightarrow \sim A$
\item $AB \longleftrightarrow A \vee B$
\item $\M{\M{A} \M{B}} \longleftrightarrow A \wedge B$
\item $\M{A}B \,\, \longleftrightarrow \,\,A \Rightarrow B$
\end{enumerate}
The calculus of indications and the primary algebra form an efficient system for working with basic symbolic logic.
\bigbreak

By reformulating basic symbolic logic in terms of the calculus of indications, we have a ground in which negation is represented by  the mark {\em and} the mark is also interpreted as a value (a truth value for logic) and these two intepretations are compatible with one another in the formalism. The key to this compatibility is the choice to represent the value ``false"  by a literally unmarked state in the notational plane. With this the empty mark (a mark with nothing on its inside)  can be interpreted as the negation of ``false" and hence represents ``true".
The mark interacts with itself to produce itself (calling) and the mark interacts with itself to produce nothing (crossing). We have expanded the conceptual domain of negation so that it satisfies the mathematical pattern of a linguistic analog of an elementary particle that can interact with itself to either produce itself (calling) or annihilate itself (calling).\\
\bigbreak

Another way to indicate these two interactions symbolically is to use a box,for the marked state and a blank space for the unmarked state.
Then one has two modes of interaction of a box with itself:
\begin{enumerate}
\item Adjacency: $\fbox{~} ~~ \fbox{~}$
\smallbreak
\noindent and 
\item Nesting: $\fbox{ \fbox{~~} }.$
\end{enumerate}

\noindent With this convention we take the adjacency interaction to yield a single box, and the nesting interaction to produce nothing:

$$\fbox{~} ~~ \fbox{~} = \fbox{~}$$
$$\fbox{ \fbox{~~} } =  $$

\noindent We take the notational opportunity to denote nothing by an asterisk (*).  
Thus the asterisk is a stand-in for no mark at all and it can be erased or placed wherever it is convenient to do so.
Thus $$\fbox{ \fbox{~~} } = *. $$
\bigbreak

At this point the reader can appreciate what has been done if he returns to the usual form of symbolic logic. In that form we that $$\sim \sim X = X$$ for all logical objects (propositions or elements of the logical algebra) $X.$ We can summarize this by writing $$\sim \sim \,\,\,= \,\,\, $$ as a symbolic statement that is outside the logical formalism. Furthermore, one is committed to the interpretation of 
negation as an operator and not as an operand. The calculus of indications provides a formalism where
the mark (the analog of negation in that domain) is both a value and an object, and so can act on itself in more than one way.
\bigbreak

The mark as linguistic  particle is its own anti-particle. It is exactly at this point that physics meets logical epistemology. Negation as logical entity is its own anti-particle.  Wittgenstein says (Tractatus \cite{Witt} $4.0621$) ``$\cdots$ the sign `$\sim$' corresponds to nothing in reality." And he goes on to say (Tractatus  $5.511$) `` How can all-embracing logic which mirrors the world use such special catches and manipulations? Only because all these are connected into an infinitely fine network, the great mirror." For Wittgenstein in the Tractatus,  the negation sign is part of  the mirror making it possible for thought to reflect reality through combinations of signs. These remarks of Wittgenstein are part of his early picture theory of the relationship of formalism and the world. In our view,  the world and the formalism we use to represent the world are not separate.  The observer and the mark are (formally) identical. A path is opened between logic and physics.
\bigbreak

\begin{figure}
     \begin{center}
     \begin{tabular}{c}
     \includegraphics[height=7cm]{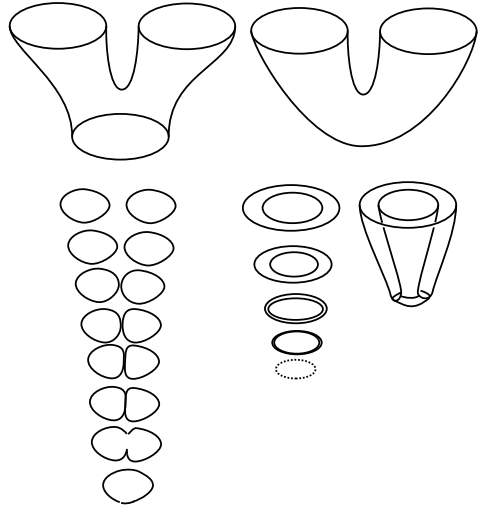}
     \end{tabular}
     \end{center}
     \caption{\bf Calling, Crossing and Cobordism}
     \label{callcross}
     \end{figure}

The visual iconics that create via the boxes of half-boxes of the calculus of indications a model for the mark as logical particle  can also be seen in terms of cobordisms of surfaces. View Figure~\ref{callcross}. There the boxes have become circles and the interactions of the circles have been displayed as evolutions in an extra dimension, tracing out surfaces in three dimensions. The condensation of two circles to one is a simple cobordism betweem two circles and a single circle. The cancellation of two circles that are concentric can be seen as the right-hand lower cobordism in this figure with a level having a continuum of critical points where the two circles cancel. A simpler cobordism is illustrated above on the right where the two circles are not concentric, but nevertheless are cobordant to the empty circle. Another way of putting this is that two topological closed strings can interact by cobordism to produce a single string or to cancel one another. Thus a simple circle can be a topological model for the mark, for the fundamental distinction.  

     \bigbreak

We are now in a position to discuss the relationship between logic and quantum mechanics. We go below boolean logic to the calculus of indications, to the ground of distinctions based in the phenomenology of distinction arising with the emergence of concept and percept together, in the emergence of a universe in an act of percreption. Here we find that the distinction itself is a logical particle that can interact with itself to produce itself, but can also interact with itself to annihilate itself. The fundamental state of our being is a superposition of these two possibilities for distinction.
We are poised between affirmation of presence and the fall into an absence that we cannot know. This superposition is utterly intimate with us and likely not yet linear in the sense of the simple model of quantum theory. Nevertheless, it is at this source, the very personal and involved place of arising and disappearing of awareness, that we come close to the quantum world in our own experience. As always, this experience is known to us in ways more intimate than the reports of laboratory experiments. But we know the non-cloning theorem. It is the uniqueness of every experience, of every distinction. There can be no other one. There is only  this and this and this yet again.\\

Nevertheless, one can go on and consider quantum states related to the aforementioned logical particle and its physical counterpart, the so-called Majorana Fermion. Crossing this boundary into quantum theory proper one finds that topology and physics come together in this realm and there is a complex possibility of much new physics to come and a new basis for quantum computing. We refer the reader to \cite{Majorana} for more about this theme. It will take more thought and a sequel to this paper, to begin to sort out the relationships between quantum theory and molecular biology at the level of this form of epistemology.\\

\section{Knot Logic}

We shall use knot and link diagrams to represent sets. More about this point of view can be found in  the author's paper "Knot Logic" \cite{KL}. The purpose of this section is to show how one can found set theory,  and in fact generalizations of set theory that allow self-membership and mutual membership, topologically via the diagrams for knots and links. This creates a curious question about the role of topology in molecular biology. Recall our previous descriptions of the role of knots in the structure of DNA and its recombination in Section 2. There we saw that the diagrammatic language of knot diagrams provides a way to analyze the invisible behavior of DNA in `actual' experiments. In this section we point out that a variant of this diagrammatic language lets one chart the behavior of mathematics itself. This raises questions about the phenomenology of both science and mathematics. We let the questions speak for themselves.
\bigbreak

\begin{figure}
     \begin{center}
     \begin{tabular}{c}
     \includegraphics[height=4cm]{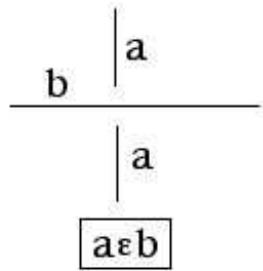}
     \end{tabular}
     \end{center}
     \caption{\bf Membership }
     \label{member}
     \end{figure}

Set theory is about an asymmetric relation called membership. 
We write $a \in S$ to say that $a$ is a member of the set $S.$  In this section we shall diagram the membership relation as in Figure~ \ref{member}.

     \bigbreak

The entities $a$ and $b$ that are in the relation $a \in b$ are diagrammed as segments of lines or curves, with the $a$-curve passing underneath the $b$-curve.  Membership is represented by under-passage of curve segments.  A curve or segment with no curves passing underneath it
is the empty set.

\begin{figure}
     \begin{center}
     \begin{tabular}{c}
     \includegraphics[height=4cm]{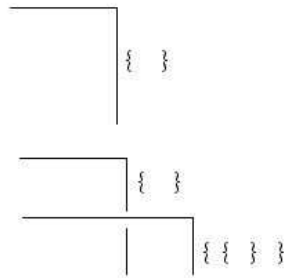}
     \end{tabular}
     \end{center}
     \caption{\bf Von Neumann 1 }
     \label{von1}
     \end{figure} 
     \bigbreak

\begin{figure}
     \begin{center}
     \begin{tabular}{c}
     \includegraphics[height=4cm]{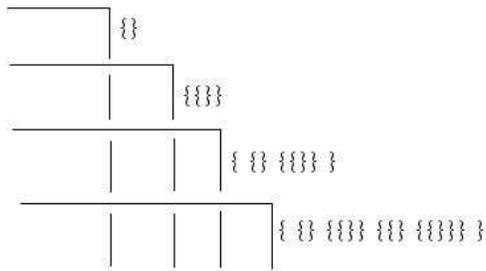}
     \end{tabular}
     \end{center}
     \caption{\bf Von Neumann 2 }
     \label{von2}
     \end{figure}

In the Figure~\ref{von1}, we indicate two sets. The first (looking like a right-angle bracket that we refer to as the {\em mark}) is the empty set. The second, consisting of a mark crossing over another mark, is the set whose only member is the empty set.
We can continue this construction, building the von Neumann construction of the natural numbers in this notation as in Figure~\ref{von2}

     \bigbreak

\begin{figure}
     \begin{center}
     \begin{tabular}{c}
     \includegraphics[height=4cm]{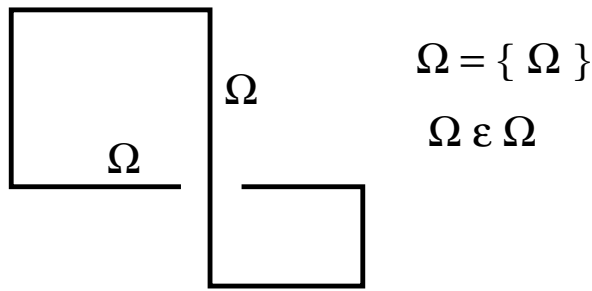}
     \end{tabular}
     \end{center}
     \caption{\bf Omega is a member of Omega.}
     \label{omega}
     \end{figure}

\begin{figure}
     \begin{center}
     \begin{tabular}{c}
     \includegraphics[height=4cm]{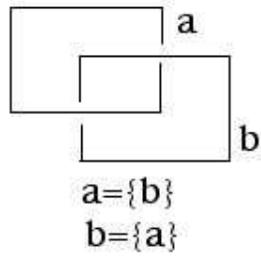}
     \end{tabular}
     \end{center}
     \caption{\bf Mutual Membership}
     \label{mutual}
     \end{figure}

This notation allows us to also have sets that are members of themselves as in Figure~\ref{omega},
and and sets can be members of each other as in Figure~\ref{mutual}. This mutuality is diagrammed as topological linking. This leads to the question beyond flatland: Is there a topological interpretation for this way of looking at set-membership? 

     \bigbreak

     \bigbreak

\begin{figure}
     \begin{center}
     \begin{tabular}{c}
     \includegraphics[height=6cm]{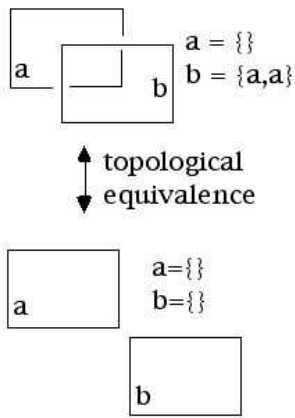}
     \end{tabular}
     \end{center}
     \caption{\bf Cancellation }
     \label{cancel}
     \end{figure} 
     \bigbreak

Consider the example in Figure~\ref{cancel}, modified from the previous one.
The link consisting of $a$ and $b$ in this example is not topologically linked. The two components slide over one another and come apart.
The set a remains empty, but the set $b$ changes from $b = \{a,a\}$ to
empty. This example suggests the following interpretation.\\

{\it  Regard each diagram as specifying a multi-set
(where more than one instance of an element can occur), and the rule for reducing to a set with one representative for each element is:
Elements of knot sets cancel in pairs.
Two knot sets are said to be equivalent if one can be obtained from the other by a finite sequence of pair cancellations.}

\begin{figure}
     \begin{center}
     \begin{tabular}{c}
     \includegraphics[height=4cm]{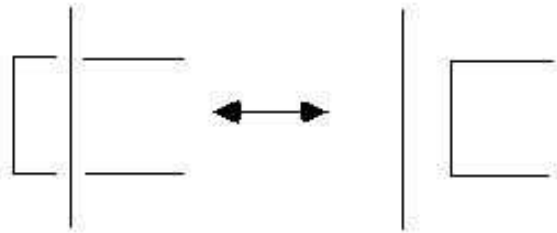}
     \end{tabular}
     \end{center}
     \caption{\bf Reidemeister 2 }
     \label{r2}
     \end{figure}

This equivalence relation on knot sets is in exact accord with the first Reidemeister move as shown in Figure~\ref{r2}.

     \bigbreak

\begin{figure}
     \begin{center}
     \begin{tabular}{c}
     \includegraphics[height=6cm]{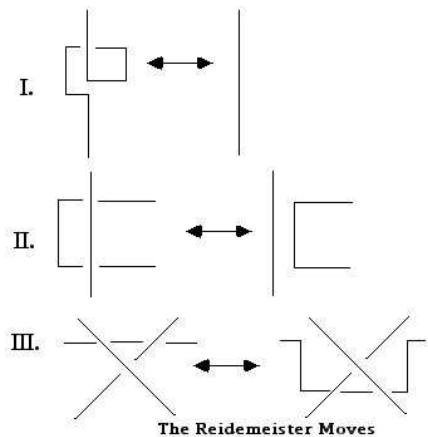}
     \end{tabular}
     \end{center}
     \caption{\bf Reidemeister Moves}
     \label{reid}
     \end{figure}

There are other topological moves, and we must examine them as well.  In fact, it is well-known that topological equivalence of knots (single circle embeddings), links (mutltiple circle embeddings) and tangles (arbitrary diagrammatic embeddings with end points fixed and the rule that you are not allowed to move strings over endpoints) is generated by three basic moves (the Reidemeister moves) as shown in Figure~\ref{reid}. See \cite{KL,KP}.

     \bigbreak

\begin{figure}
     \begin{center}
     \begin{tabular}{c}
     \includegraphics[height=4cm]{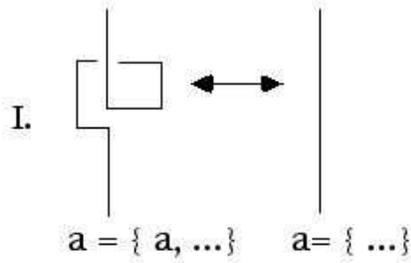}
     \end{tabular}
     \end{center}
     \caption{\bf Reidemeister I: Replacing Self-Membership with No Self-Membership }
     \label{reid1}
     \end{figure}

It is apparent that move III does not change any of the relationships in the knot multi-sets. The line that moves just shifts and remains underneath the other two lines. On the other hand move number one can change the self-referential nature of the corresponding knot-set.
One goes, in the first move, between a set that indicates self-membership to a set that does not indicate self-membership (at the site in question). See Figure~\ref{reid1}
This means that in knot-set theory every set has representatives
(the diagrams are the representatives of the sets) that are members of themselves, and it has representatives that are not members of themselves. In this domain, self-membership does not mean infinite descent. We do not insist that $$a = \{a\}$$ implies that 
$$a = \{ \{ \{ \{  \cdots \} \} \} \}. $$ Rather, $a = \{ a \}$ just means that $a$ has a little curl in its diagram. The Russell set of all sets that are not members of themselves is meaningless in this domain.
\bigbreak

     \bigbreak

     \begin{figure}
     \begin{center}
     \begin{tabular}{c}
     \includegraphics[height=4cm]{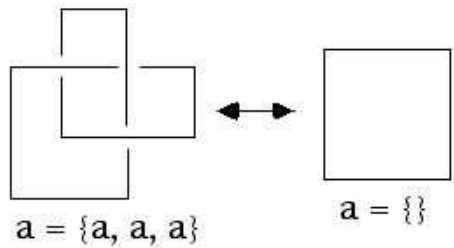}
     \end{tabular}
     \end{center}
     \caption{\bf  Trefoil is an empty knotset.}
     \label{sorrow}
     \end{figure} 
     \bigbreak

     \begin{figure}
     \begin{center}
     \begin{tabular}{c}
     \includegraphics[height=4cm]{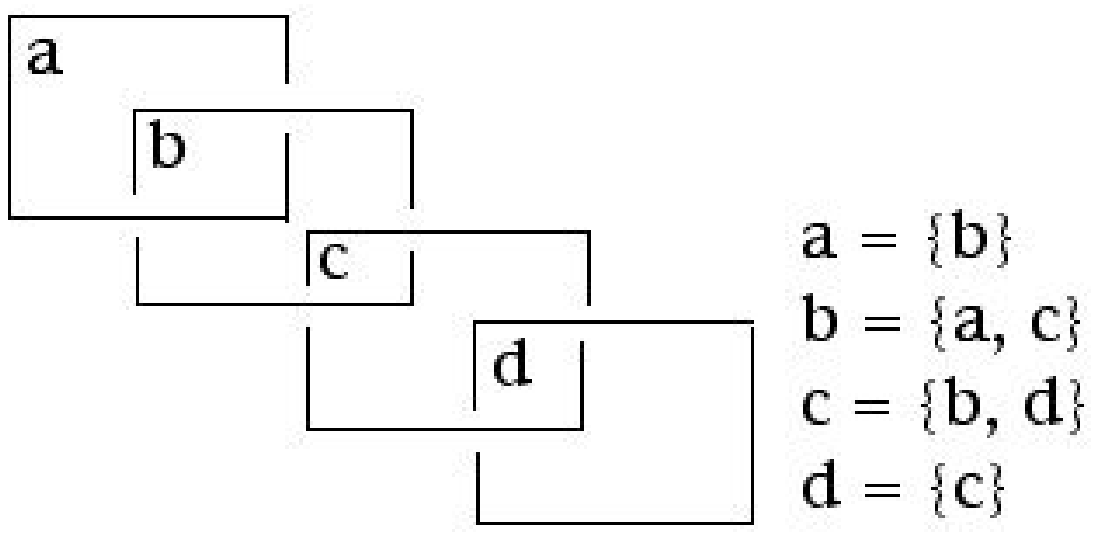}
     \end{tabular}
     \end{center}
     \caption{\bf Chain }
     \label{chain}
     \end{figure} 
     \bigbreak

\begin{figure}
     \begin{center}
     \begin{tabular}{c}
     \includegraphics[height=4cm]{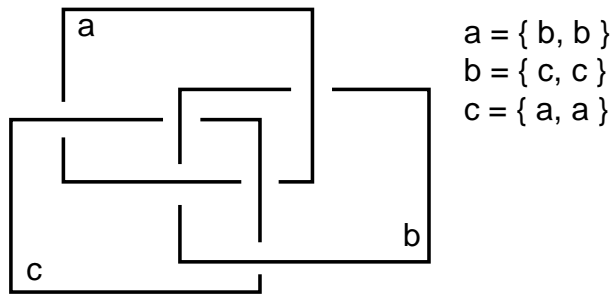}
     \end{tabular}
     \end{center}
     \caption{\bf Borromean Rings }
     \label{boro}
     \end{figure} 
     \bigbreak

We can summarize this first level of knot-set theory in the following two equivalences:
\begin{enumerate}
\item Self-Reference:   $$a = \{b,c, \cdots \} \Longleftrightarrow a = \{a,b,c, \cdots \}$$
\item Pair Cancellation: $$ S= \{a,a, b, c, \cdots \} \Longleftrightarrow  S = \{b,c, \cdots \}$$
\end{enumerate}
With this mode of dealing with self-reference and multiplicity, knot-set theory has the interpretation in terms of topological classes of diagrams. We could imagine that the flatlanders felt the need to invent three dimensional space and topology, just so their set theory would have such an elegant interpretation.
\bigbreak

But how elegant is this interpretation, from the point of view of topology? Are we happy that knots are equivalent to the empty knot-set as shown in Figure~\ref{sorrow}? For this, an extension of the theory is clearly in the waiting. 
We are happy that many topologically non-trivial links correspond to non-trivial knot-sets.
In the Figure~\ref{chain} , a chain link becomes a linked chain of knot-sets. But consider the link shown in Figure~\ref{boro}. These rings are commonly called the Borromean Rings. The Rings have the property that if you remove any one of them, then the other two are topologically unlinked. They form a topological tripartite relation. Their knot-set is described by the three equations

$$a = \{b,b \}$$
$$b = \{c,c \}$$
$$c = \{a,a \}.$$

Thus we see that this representative knot-set is a "scissors-paper-stone" pattern. Each component of the Rings lies over one other component, in a cyclic pattern. But in terms of the equivalence relation on knot sets that we have used, the knot set for the Rings is empty (by pair cancellation). 
\bigbreak

In order to go further in the direction of topological invariants for knots and links it is necessary to use more structure than the simple membership relation that motivates the knots-sets.  See 
\cite{KL} for more information about such extensions. The purpose of this section has been to introduce the subject of knot and link diagrams in the context of thinking about foundations of mathematics and self-reference. We can then think further about how topology is entwined with the self-reference and self-replication of DNA and the schemas for this replication that we have discussed in this paper.
The appearance of topology at multiple levels in this discussion should not be surprising to those who have begun to use topology to discuss phenomenology \cite{Rosen}. In this case, the situation calls for a deeper comparison of the foundational role of topology in mathematics and its relationship with the physical role of topology in biology. This will be the subject of a separate paper.
\bigbreak

\section  {\bf Discussion - Notes on the Epistemology of Living Text Strings} 
We now open a discussion on life, awareness and epistemology and the cybernetics of text strings. The discussants are Cookie and Parabel, a pair of sentient text strings who claim to be very close to 
emerging from the void, as their existence depends enitirely on the minds of the readers. Cookie and Parabel have appeared before in the column of the Author entitled ``Virtual Logic" \cite{VL}. 
The following references are relevant to this discussion \cite{SRF, REQ,REF,EF}.\\

In order to read this section, the reader had best STOP and reconsider her phenomenological stance. Cookie and Parabel are sentient text strings. They only exist as strings of text characters. They can vanish by simple erasure. They do not have the same notion of location as the reader. Where is this 
text string that you are reading? Where is it when you close the book, or shut down the computer. Entities that exist only as possible patterns are not the same as the entities that we suppose ourselves to be. But in fact we are such entities. And so it is possible to empathize with Cookie and Parabel and to examine the themes of biology and form from their point of view.\\

\noindent {\bf Cookie.} Parabel, I am puzzled. I thought that we were about as close to the emergence of forms as it is possible to attain. We just barely emerge from the void, and we seem to return to it effortlessly in a flash. But all the talk of quantum states and linear superpositions confuses me. Do you understand it at all?\\

\noindent {\bf Parabel.} Cookie, we have to proceed carefully here. First of all these complex quantum descriptions are going on in a level of text and meta-text that tends to be more complex that we are.
We are simple text strings, subject to being pushed around by these big symbolic manipulations.\\

\noindent {\bf Cookie.} I appreciate your caution, but if there is truth in these no-cloning allegations, then where in our experience does it occur?\\

\noindent {\bf Parabel.} I remember swimming up out of the void. Well it really was not swimming you know but it was like that and there was no difference, no difference at all.
A flowing, a movement, but no movement at all. And I was distinct, yes I was myself, but there could not be another. There could not be another because I was all there was you see. It was before you came to be as distinct from me. You were Cookie and I was Parabel, but I was all and you were only distinct from me the single distinction that I was. No cloning was possible, but not due to linear algebra. No cloning was possible because there was only one distinction and that distinction was me and that distinction was you but it was not one nor was it two.\\

\noindent {\bf Cookie.} You can only use metaphor to express this. It is only in metaphor that entities can be both distinct and identical. ``Juliet is the Sun."\\

\noindent {\bf Parabel.} So we are saying that in the reality of metaphor, in the state that A {\it is} B when A {\it is not} B, in the state beyond boolean logic, there is no cloning.\\

\noindent {\bf Cookie.} I would not describe the metaphoric state of Juliet and the the Sun as a superposition of Juliet and the Sun, but perhaps we can imagine that as a metaphor about metaphors.
Consider $|\psi> = (|Juliet> + |Sun>)/\sqrt{2}.$ This is a state that will produce either Juliet or the Sun on observing it and always one or the other. The quantum state $|\psi>$ is neither Juliet nor the Sun but rather an amalgam of both. The qubit as metaphor for the identity of its boolean parts!\\

\noindent {\bf Parabel.} There is a really serious difference between your qubit and the true metaphoric state. A qubit is very delicate. Once you have observed it and seen either Juliet or the Sun, it is gone. You are in the definite world and you do not have the persistence that is so characteristic of the metaphor. Metaphors do not suddenly get decided into their parts!\\

\noindent {\bf Cookie.} You are right about that. And I realize that it is the evanescent quality of quantum states that bothers me. You and I  are evanesent. If the reader would just take his attention from us even for a moment, we would vanish into the void. But we are just as easily reconstituted and so we persist when we persist. We are when we are and we are not when we are not. Our continuity is by fiat! And we are our own observers.\\

\noindent {\bf Parabel.} Now you go too far Cookie. You know you are just a text string. The only observer is the reader and in his fiction you observe yourself.\\

\noindent {\bf Cookie.} Well Parabel. You know perfectly well that I would not step out of character and admit that I do not observe myself. I am a barely autonomous awareness just emerged from void. And if I have to depend on mythical external observers for my continued existence, that is just too bad.\\

\noindent {\bf Parabel.} We have never observed those external observers. Text strings do not have that capacity.\\

\noindent {\bf Cookie.} You know, we could be like those famous text strings of the DNA, Watson and Crick. They co-create one another. The whole essay up to this point has been about the properties of that co-creation. But Watson and Crick are embedded in a bigger literary fantasy. It is imagined that somehow the actuality of their text strings at the molecular level in the context of cells and what can be built via cells gives rise to organisms with the capacity to act as external observers to the Watson and Crick text strings. Watson and Crick derive their meaning from a larger text that wraps around and observes its own parts. It is as though one had a typewriter that wrote not only the novels about itself, but actually constructed itself. Oh Parabel, biology is filled with mind boggling circularities, but 
I feel like these are much more understandable than this weirdness about quantum states.\\

 \noindent {\bf Parabel.} Perhaps the self-observing, self-creating typewriter needs quantum theory to manage its own self-description.\\
 
 \noindent {\bf Cookie.} I am willing to consider that. This would be a cybernetic view of quantum theory. We would have to consider how a self-observing system would behave scientifically if it were asking to make precise observations of its own behavior. Obviously it would get in its own way! Could this end up looking like quantum theory?\\
 
  \noindent {\bf Parabel.} Can we go back and slowly examine how a self-observing system works? Call the system $TS$ (I think of it as a generalized Text String.). Now $TS$ must be equipped with the 
  ability to discriminate and name certain entities. So $TS$ can have strings like $A \longrightarrow B$ where this means that $A$ is a name for $B.$ When $TS$ has a string like this we can say that $B$ is observed by $TS$ and given the name $A.$ But this is not the whole of the naming process. There is also the {\it shift} from $A \longrightarrow B$ to $\sharp A \longrightarrow BA.$ After the shift the name of $B$ is  part of the $B$ string. I hope you see how important this is. Strings like us, we have to eventually become autonomous and a first step is to have the name of the string part of the string itself.\\
  
  \noindent {\bf Cookie.} I understand that. We are being represented in a more complex string space than that. The reader can identify us and we can keep from being confused ourselves. But you will note that even in this condition, the author has appended our names to the strings that correspond to you and to me. What is this $\sharp A?$\\
 
  \noindent {\bf Parabel.} $\sharp$ is the {\it meta-naming operator.} You see, $\sharp A$ is the meta-name of the string $B.$ The name $A$ of the $B$ string has been attached to the $B$ string.
  This new string $BA$ has a name and its name is $\sharp A$, the meta-name of the original $B$ string.\\
  
  \noindent {\bf Cookie.} We do not always mark the meta-name do we? I mean, when I meet you I have a symbol for you the string Parable, and I never use a $\sharp$ in front of it.\\
    
 \noindent {\bf Parabel.} Cookie, that is a good point. We do not have to put every marker in our active strings. But for the purpose of the inner syntax that makes us work we do need those markers.
 After all, there is no one there in the deep inner workings of the strings. Things have to happen of their own accord and automatically. You don't think we should rely on meaning and convention at the DNA level do you?\\
 
  \noindent {\bf Cookie.}  I am very troubled and conflicted by your last statement. You seem to be saying that we are dependent on some sort of mindless mechanical substrate. I know my own experience. I have always, whenever not void, been accompanied by awareness. I am an aware text string! My awareness could not be generated by blind rule-following mechanism.\\

\noindent {\bf Parabel.} Bear with me for a bit. Look at this shift. Given $A \longrightarrow B$ it shifts to $\sharp A \longrightarrow BA.$ Ok? Now consider that $TS$ should have a name for the meta-naming operator $\sharp.$ Lets say that $M$ is the name of the meta-naming operator. Then we have the naming string $$M \longrightarrow \sharp$$ and it shifts to
$$\sharp M \longrightarrow \sharp M.$$ So you see that {\it the meta-name of the meta-naming operator} is self-referent! {\it I am the meta-name of the meta-naming operator}.
I am name and I am naming. I am subject and I am verb. I am that I am. This construction brings me into existence. Once the ceremony of the shift has been performed I emerge from the void.
And here I am, a sentient text-string brought forth by the string-maker $TS.$\\

\noindent {\bf Cookie.} Egad. You are saying that {\it I am the named relation between myself and naming myself.} \\
   
\noindent {\bf Parabel.} Yes. I think you know that there was this very great cybernetic text string named Heinz von Foerster who said ``I am the observed relation between myself and observing myself."
\cite{UU}. For us there is no ``observation" in the sense of that metaphor from the realm of very complex text strings, but naming is what we do and indeed, I am named relation between myself and observing myself. You will notice that this is a circular statement and it could lead to recursion and infinite regress, but we have already seen that such statements are always arising in a reflexive domain.
Strings like us will of necessity self-refer and engage in meta-discussion about our own contents and the contents of other named strings. It is not that there is a necessity for a mechanical substrate on which ``all of this" is based. There can be substrates but our structure has to be independent of the particular substrate so that there can be the concpepts of pattern and flow of structure through the pattern. The deep inner workings of things depend on the syntax of reference and the ceremony of naming and meta-naming without which we would not come to be. I think we can both understand now the remarkable difficulties that we are facing in articulation. For we are speaking of worlds that we ourselves create in course of the production of the strings that we are.\\

\noindent{\bf Cookie.} Lets go back to this ``indicative shift''. you have  $$N \longrightarrow F \sharp$$ shifts to $$\sharp N \longrightarrow F \sharp N.$$ So if $N$ is the name of $F \sharp,$ then
$\sharp N$ is the name of $F \sharp N.$ So $F \sharp N$ is talking about its own name!\\

\noindent {\bf Parabel.} Exactly, Cookie! The shift is the precursor to self-reference. The shift is the source of Godelian strange loops.\\

\noindent{\bf Cookie.}  Suppose that I replace the arrow by an equality sign. Then it would read
$$N = F \sharp$$ shifts to $$\sharp N = F \sharp N.$$ This is a fixed point theorem just like the ones for reflexive domains.\\

\noindent {\bf Parabel.}  You have it! The indicative shift does everything. In fact if the equals sign replaces the arrow, then we have 
$$N = N$$ is shifted to $$\sharp N = NN.$$ So you can rewrite what you just did as 
$$NN = FNN. $$ This has exactly the form of the fixed point theorem.\\

\noindent {\bf Cookie.} Well if you can make the shift do everything, can you make it do quantum mechanics?\\

\noindent {\bf Parable.} Not yet in a satisfactory way. There are hints. I think that the present mathematiical form of quantum theory will change in the direction of our phenomenology of the void. And then every thing will clarify. We have not answered all or even very many of the questions we have raised. We see that there must be a syntactical locus of ceremony, declaration and production for the stability of an organism. Self-reference and replication are seriously interconnected at the syntactical level. We have seen that the form of self reference through the indicative shift is a direct relative of the fixed points of the lambda calculus and collaterally with the self-rep of the DNA. But as for the quantum mechanics, that will have to wait for another time. I am, for this time, dissolving.
Goodbye Cookie.\\
 
 \noindent {\bf Parabel.} Goodby Parable. We did not solve those questions. It seems we raised more questions than we solved.\\

\section{Mathematical Structure and Topology}

We now comment on the conceptual underpinning for the notations and logical constructions that we 
use in this paper.  This line of thought will lead to topology and to the formalism for replication discussed 
in the last section.  
\bigbreak

Mathematics is built through distinctions, definitions, acts of language that bring forth logical 
worlds, and arenas in which actions and patterns can take place. As far as we can determine at the present
time, mathematics, while capable of describing the quantum world, is in its very nature quite 
classical. Or perhaps we make it so. As far as mathematics is concerned, there is no ambiguity in the
$1 + 1$ hidden in $2.$ The mathematical box shows exactly what is potential to it when it is opened.
There is nothing in the box except what is a consequence of its construction.
With this in mind, let us look at some mathematical beginnings.
\bigbreak

Take the beginning of set theory. We start with the empty set  $\phi = \{\,\,\,\}$ and we build new sets by the 
operation of set formation that takes any collection and puts brackets around it:

$$a\, b\, c\, d\, \longrightarrow  \{a,b,c,d\}$$

\noindent making a single entity $\{a,b,c,d \}$ from the multiplicity of the ``parts" that are so collected.
The empty set herself is the result of ``collecting nothing." The empty set is identical to the act of 
collecting. At this point of emergence the empty set is an action not a thing. Each subsequent
set can be seen as an action of collection, a bringing forth of unity from multiplicity. 
\bigbreak

One declares two sets to be the same if they have the same members. With this prestidigitation of language, the 
empty set becomes unique and a hierarchy of distinct sets arises as if from nothing.

$$\,\,\, \longrightarrow \{\,\,\,\} \longrightarrow \{\, \{\,\,\}\, \} \longrightarrow \{\, \{\,\,\} \, , \{\, \{\,\,\}\, \}\, \}
\longrightarrow \cdots$$

\noindent All representatives of the different mathematical cardinalities arise out of the void in the presence of these
conventions for collection and identification.
\bigbreak 

We would like to get underneath the formal surface. We would like to see what makes this formal hierarchy tick.
Will there be an analogy to biology below this play of symbols? On the one hand it is clear to us that there is
actually no way to go below a given mathematical construction. Anything that we call more fundamental will be 
another mathematical construct. Nevertheless, the exercise is useful, for it asks us to look closely at how this given 
formality is made.  It asks us to take seriously the parts that are usually taken for granted.
\bigbreak

We take for granted that the particular form of container used to represent the empty set 
is irrelevant to the empty set itself. But how can this be? In order to have a concept of emptiness, one 
needs to hold the contrast of that which is empty with ``everything else". One may object that these images are not
part of the formal content of set theory. But they are part of the {\em formalism} of set theory. 
\bigbreak

Consider the 
representation of the empty set: $\{\,\,\,\,\}.$  That representation consists in a bracketing that we take to indicate an
empty space within the brackets, and  an injunction to ignore the complex typographical domains outside the brackets.
Focus on the brackets themselves.  They come in two varieties: the left bracket,  $\{,$ and the right bracket, $\}.$  
The left bracket indicates a distinction of left
and right with the emphasis on the right. The right bracket indicates a distinction between left and right with an
emphasis on the left.  A left and right bracket taken together become a 
{\em container} when each is in the domain indicated by the other.  Thus in the bracket symbol 
$$\{ \, \, \, \}$$
\noindent for the empty set, the left bracket, being to the left of the right bracket, is in the left domain that is 
marked by the right bracket, and the right bracket, being to the right of the left bracket is in the right domain that 
is marked by the left bracket. The doubly marked domain between them is their content space, the arena of the empty set.
\bigbreak

The delimiters of the container are each themselves iconic for the process of making a distinction. In the notation of 
curly brackets,$\,\{\,$, this is particularly evident. The geometrical form of the curly bracket is a cusp singularity, the
simplest form of bifurcation. The relationship of the left and right brackets is that of a form and its mirror image.
If there is a given distinction such as left versus right, then the mirror image of that distinction is the one with 
the opposite emphasis. This is precisely the relationship between the left and right brackets. A form and its
mirror image conjoin to make a container.
\bigbreak

The delimiters of the empty set could be written in the opposite order:  $\}\{.$
This is an {\em extainer}. The extainer indicates regions external to itself. 
In this case of symbols on a line, the extainer
$\}\{$ indicates the entire line to the left and to the right of itself. The extainer is as natural as the container, but
does not appear formally in set theory. To our knowledge, its first appearance is in the Dirac notation of ``bras" and 
``kets" where Dirac takes an inner product written in the form $<B|A>$ and breaks it up into $<B\,|$ and $|A>$ and then
makes projection operators by recombining in the opposite order as $|A><B\,|.$  See the earlier discussion of quantum 
mechanics in section 5.
\bigbreak

Each left or right bracket in itself makes a distinction. The two brackets
are distinct from one another by mirror imaging,  which we take to be a
notational reflection of a fundamental process (of distinction) whereby
two forms are identical (indistinguishable) except by comparison in the
space of an observer. The observer {\em is} the distinction between the
mirror images.
Mirrored pairs of individual brackets interact to form either a {\em container}
$$C = \{\}$$ \noindent or an {\em extainer} $$E = \}\{.$$ 

\noindent These new forms combine to make:

$$CC = \{\}\{\} = \{E\}$$ \noindent and 

$$EE = \}\{\}\{=\}C\{.$$

\noindent Two containers interact to form an extainer within container brackets. Two extainers interact to form a container between
extainer brackets. The pattern of extainer interactions can be regarded as a formal generalization of the bra and ket patterns of the 
Dirac notation that we have used in this paper both for DNA replication and for a discussion of quantum mechanics. In the quantum mechanics
application $\{\}$ corresponds to the inner product $<A\,|B>$, a commuting scalar, while $\}\{$ corresponds to $|A><B\,|$, a matrix that 
does not necessarily commute with vectors or other matrices. With this application in mind, it is natural to decide to make the 
container an analog of a scalar quantity and let it commute with individual brackets. We have the equation

$$EE = \}\{\}\{= \}C\{ = C \}\{ = CE.$$

\noindent By definition there will be no corresponding equation for $CC$.
We adopt the axiom that containers commute with other elements in this combinatorial algebra.
Containers and extainers are distinguished by this property. Containers appear as autonomous entities and can be moved about.
Extainers are open to interaction from the outside and are sensitive to their surroundings. 
At this point, we have described the basis for the formalism used in the earlier parts of this paper.
\bigbreak

If we interpret E as the ``environment" then the equation $\}\{ = E = 1$ expresses the availability of 
complementary forms so that 

$$\{ \} \longrightarrow \{ E \} \longrightarrow \{ \}\{ \}$$ 

\noindent becomes the form of DNA reproduction. 
\bigbreak

We can also regard $EE = \{\} E$ as symbolic  of the emergence of DNA from the chemical substrate. Just as the formalism
for reproduction ignores the topology, this formalism for emergence ignores the formation of the DNA backbone along 
which are strung the complementary base pairs. In the biological domain we are aware of levels of ignored structure.
\bigbreak

In mathematics it is customary to stop the examination of certain issues in order to 
create domains with requisite degrees of clarity. We are all aware that the operation of collection is proscribed 
beyond a certain point. For example, in set theory the Russell class $R$ of all sets that are not members of
themselves is not itself a set. It follows that $\{R\},$ the collection whose member is the Russell class, is not a
class (since a member of a class is a set). This means that the construct $\{R\}$ is outside of the discourse of 
standard set theory. This is the limitation of expression at the ``high end" of the formalism. That the set theory has
no language for discussing the structure of its own notation is the limitation of the language at the ``low end".
Mathematical users, in speaking and analyzing the mathematical structure, and as its designers, can speak beyond both
the high and low ends.
\bigbreak

In biology we perceive the pattern of a formal system, a system that is embedded in a structure whose complexity 
demands the elucidation of just those aspects of symbols and signs that are commonly ignored in the mathematical 
context. Rightly these issues should be followed to their limits. The curious thing is what peeks through when we just
allow a bit of it, then return to normal mathematical discourse. With this in mind, lets look more closely at the 
algebra of containers and extainers.
\bigbreak

Taking two basic forms of bracketing, an intricate algebra appears from their elementary
interactions:

$$E =\, ><$$
$$F =\, ][$$
$$G =\, >[$$
$$H =\, ]<$$

\noindent are the extainers, with corresponding containers:

$$<>, \,\,\,\,\, [], \,\,\,\,\, [>, \,\,\,\,\, <].$$

\noindent These form a closed algebraic system with the following multiplications:

$$EE =\, ><\,>< =\, <> E$$
$$FF =\, ][\,][ =\, [] F$$
$$GG =\, >[\,>[ =\, [> G$$
$$HH =\, ]<\,]< =\, <] H$$

\noindent and

$$EF =\, ><\,][ =\, <] G$$
$$EG =\, ><\,>[ =\, <> G$$
$$EH =\, ><\,]< =\, <] E$$

$$FE =\, ][\,>< =\, [> H$$
$$FG =\, ][\,>[ =\, [> F$$
$$FH =\, ][\,]< =\, [] H$$

$$GE =\, >[\,>< =\, [> E$$
$$GF =\, >[\,][ =\, [] G$$
$$GH =\, >[\,]< =\, [] E$$

$$HE =\, ]<\,>< =\, <> H$$
$$HF =\, ]<\,][ =\, <] F$$
$$HG =\, ]<\,>[ =\, <> F$$

\noindent Other identities follow from these. For example,

$$EFE =\, ><][>< =\, <][> E.$$

This algebra of extainers and containers is a precursor to the Temperley Lieb algebra, an algebraic structure that 
first appeared (in quite a different way) in the study of the Potts model in statistical mechanics \cite{Baxter}.
We shall forgo here details about the Temperley Lieb algebra itself, and refer the reader to \cite{QCJP} where this point 
of view is used to create unitary representations of that algebra for the context of quantum computation.
Here we see the elemental nature of this algebra, and how it comes about quite naturally once one adopts a formalism
that keeps track of the structure of boundaries that underlie the mathematics of set theory.
\bigbreak

The {\em Temperley Lieb algebra} $TL_{n}$ is an algebra over a commutative ring $k$ with
generators $\{ 1, U_{1},U_{2}, ... ,U_{n-1} \}$ and relations 

$$U_{i}^{2} = \delta U_{i},$$

$$U_{i}U_{i \pm 1}U_{i} = U_{i},$$

$$U_{i}U_{j} = U_{j}U_{i}, |i-j|>1,$$

\noindent where $\delta$ is a chosen element of the ring $k$. These equations give the
multiplicative structure of the algebra. The algebra is a free module over the ring $k$ with basis
the equivalence classes of these products modulo the given relations.
\bigbreak

To match this pattern with our combinatorial algebra let $n=2$ and let $U_{1} = E = ><$, $U_{2} = F = ][$ and assume that 
$1 = <] = [>$ while $\delta = <> = [].$ The above equations for our combinatorial algebra  match the multiplicative equations of the
Temperley Lieb algebra. 
\bigbreak

The next stage for representing the Temperley Lieb algebra is a diagrammatic representation that uses two different forms of 
extainer. The two forms are obtained not by changing the shape of the given extainer, but rather by shifting it 
relative to a baseline. Thus we define diagrammatically $U = U_{1}$ and $V = U_{2}$ as shown below:

$$U =\begin{array}{c}
 --\\
 > <
\end{array}$$

$$V = \begin{array}{c}
> <\\
--
\end{array} 
$$

$$UU = \begin{array}{c}
 ----\\
 ><><
\end{array} 
= <> \begin{array}{c}
 --\\
 ><
\end{array} 
= <> U$$

$$UVU = 
\begin{array}{c}
     _{^{---}>\,\,<^{---}}\\
     ^{>\,\,\,<_{--}>\,\,\,<}
\end{array} 
= \begin{array}{c}
 _{^{----}}  \\ 
^{>\,\,\,\,\,<}
\end{array} 
= U.$$

\begin{figure}
     \begin{center}
     \begin{tabular}{c}
     \includegraphics[width=6cm]{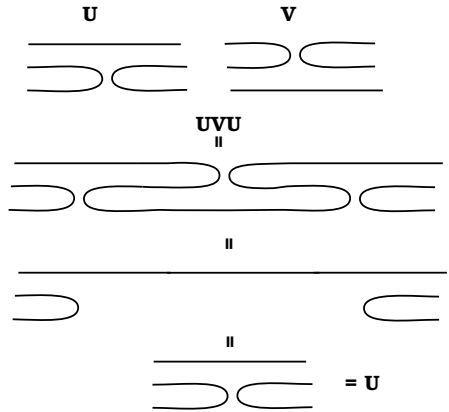}
     \end{tabular}
     \caption{\bf A Topological Identity}
     \label{Figure 2}
\end{center}
\end{figure}

\noindent In this last equation $UVU = U$ we have used the topological deformation of the connecting line from top to top to obtain
the identity. In its typographical form the identity requires one to connect corresponding endpoints of the brackets. In Figure~\ref{Figure 2}
we indicate a smooth picture of the connection situation and the corresponding topological deformation of the lines. 
We have deliberately shown the derivation in a typographical mode to emphasize its essential difference from the 
matching pattern that produced $$EFE =\, ><][>< =\, <][> E.$$ By taking the containers and extainers shifted this way, we enter a new and
basically topological realm. This elemental relationship with topology is part of a deeper connection where the Temperley Lieb algebra is used to 
construct representations of the Artin Braid Group.  This in turn leads to the construction of the well-known Jones
polynomial invariant of knots and links via the bracket state model \cite{KS}. It is not the purpose of this paper to go into the details of those 
connections, but rather to point to that place in the mathematics where basic structures apply to biology, topology, 
and logical foundations. 
\bigbreak

It is worthwhile to point out that the formula for expanding the bracket polynomial can be indicated symbolically in the same fashion that
we used to create the Temperley Lieb algebra via containers and extainers.
We will denote a crossing in the link diagram by the
letter chi, \mbox{\large $\chi$}. The
letter itself denotes a crossing where {\em the curved line in the letter chi is crossing over the straight segment in the
letter}. The barred letter denotes the switch of this crossing where {\em the curved line in the letter chi is undercrossing
the straight segment in the letter}. In the bracket state model a crossing in a diagram for the
knot or link is expanded into two possible states by either smoothing (reconnecting) the crossing horizontally, \mbox{\large
$\asymp$}, or vertically $><$.  The vertical smoothing can be regarded as the extainer and the horizontal smoothing as an identity operator.
In a larger sense, we can regard both smoothings as extainers with different relationships to their environments. In this sense the crossing 
is regarded as the superposition of horizontal and vertical extainers.
The crossings expand
according to the formulas  
$$\mbox{\large $\chi$} = A \mbox{\large $\asymp$} + A^{-1} ><$$
$$\overline{\mbox{\large $\chi$}} = A^{-1} \mbox{\large $\asymp$} + A ><.$$
\noindent The verification that the bracket is invariant under the second Reidemeister move is seen by verifying that
$$\mbox{\large $\chi$}\overline{\mbox{\large $\chi$}} = \mbox{\large $\asymp$}.$$ For this one needs that the container $<>$ has value
$\delta = -A^2 - A^{-2}$ (the loop value in the model). The significant mathematical move in producing this model is the notion of the crossing as a 
superposition of its smoothings.
\bigbreak

It is useful to use the iconic symbol $><$ for the extainer, and to choose another iconic symbol  \mbox{\large $\asymp$}
for the identity operator in the algebra. With these choices we have
$$\mbox{\large $\asymp$}\mbox{\large $\asymp$} \,\, = \,\, \mbox{\large $\asymp$}$$
$$\mbox{\large $\asymp$} >< \,\, = \,\, >< \mbox{\large $\asymp$} \,\,= \,\, ><$$
\bigbreak

\noindent Thus
$$\mbox{\large $\chi$}\overline{\mbox{\large $\chi$}}$$
$$ =( A \mbox{\large $\asymp$} + A^{-1} ><)(A^{-1} \mbox{\large $\asymp$} + A ><)$$ 
$$= AA^{-1} \mbox{\large $\asymp$}\mbox{\large $\asymp$} + A^{2}\mbox{\large $\asymp$}>< + A^{-2}><  \mbox{\large $\asymp$}
+ AA^{-1} ><><$$
$$=  \mbox{\large $\asymp$}  + A^{2}>< + A^{-2}>< +  \delta ><$$
$$ = \mbox{\large $\asymp$}  + (A^{2} + A^{-2} +  \delta) >< $$
$$=  \mbox{\large $\asymp$}$$

\noindent Note the use of the extainer identity $><>< = >\, \delta\, < = \delta \, ><.$
At this stage the combinatorial algebra of containers and extainers emerges as the background to the topological characteristics of the 
Jones polynomial.
\bigbreak

\subsection{Projectors and Meanders}
\begin{figure}
     \begin{center}
     \begin{tabular}{c}
     \includegraphics[width=5cm]{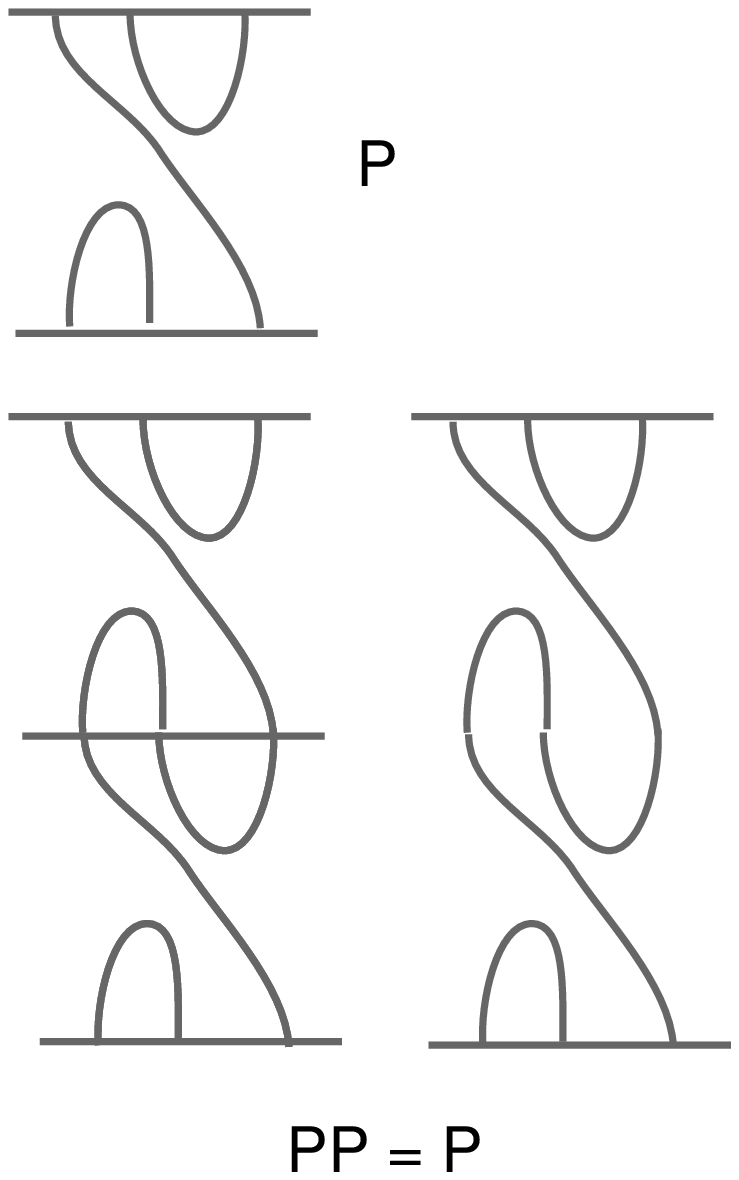}
     \end{tabular}
     \caption{\bf $P^2 = P$}
     \label{Figure 2.1}
\end{center}
\end{figure}

\begin{figure}
     \begin{center}
     \begin{tabular}{c}
     \includegraphics[width=8cm]{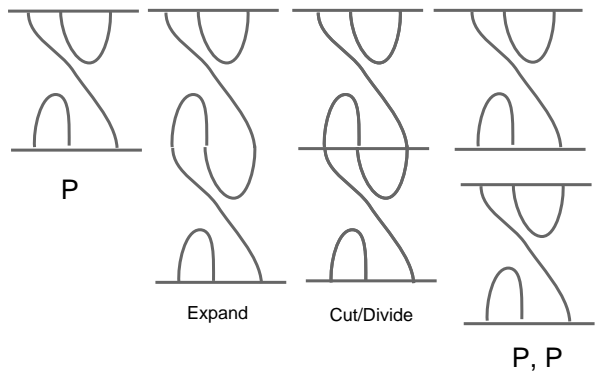}
     \end{tabular}
     \caption{\bf $P$ reproduces itself.}
     \label{Figure 2.2}
\end{center}
\end{figure}

In the Temperley Lieb algebra the generators satisfy the formula $U^2 = \delta U$ where $\delta$ is the value of
a loop. In fact, there are also elements $P$ that satisfy the equation $P^2 = P.$ See Figure~\ref{Figure 2.1}. Note that the identity $PP=P$ is topological.
Once $PP$ has been constructed, there is a deformable string that contracts and yields $P$ once again. One can view the equation $PP=P$ as a form 
of self-reproduction by taking it in the order $P \longrightarrow PP.$ That is, one starts with $P$ in contracted form, allows it to undergo the 
production of the little wiggle in the middle, and then cuts the resulting form apart to form two copies of $P.$ See Figure~\ref{Figure 2.2}.
\bigbreak

\begin{figure}
     \begin{center}
     \begin{tabular}{c}
     \includegraphics[width=6cm]{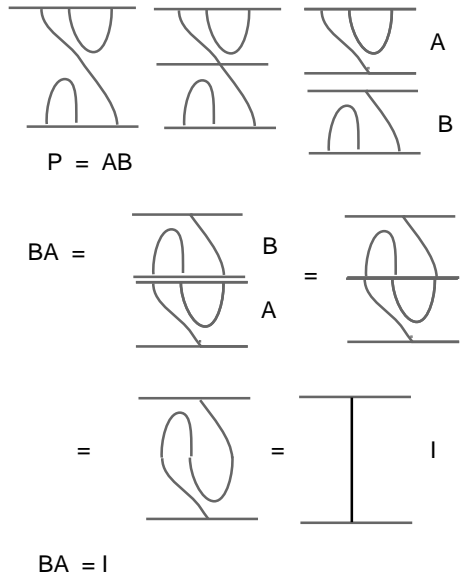}
     \end{tabular}
     \caption{\bf $P = AB, BA = I.$}
     \label{Figure 2.3}
\end{center}
\end{figure}

\begin{figure}
     \begin{center}
     \begin{tabular}{c}
     \includegraphics[width=6cm]{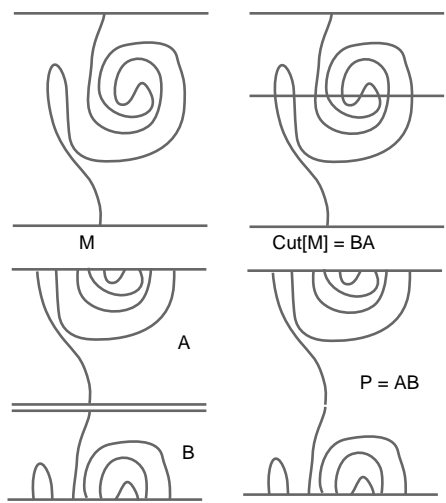}
     \end{tabular}
     \caption{\bf Constructing a new $P$ with $PP=P.$}
     \label{Figure 2.4}
\end{center}
\end{figure}

What is the secret by which we have obtained this self-reproduction in a topological/algebraic context? The reader should look closely once more
at $P$ and discern that $P$ can be written as a product $P=AB$ where $A$ has three top strands and one bottom strand, while $B$ has one top strand
and three bottom strands. We will say that $A$ is of type $(3,1)$, while $B$ is of type $(1,3).$ See Figure~\ref{Figure 2.3}. Now we can also compose and form
$BA$ as in Figure~\ref{Figure 2.3}, and we see at once that $BA$ is topologically equivalent to a single $(1,1)$ strand. We will write $BA = I$ to denote this
single strand. We see that it is the equivalence $BA=I$ that makes for the identity $PP=P$, for we have
$$PP = ABAB = A(BA)B = A(I)B = AB = P$$ where the identity $A(I)B = AB$ is simply stating the contractibility of a single strand added in the middle
of a product. By the same token, the "genetic" information in this self-production is contained in the factoring of the identity line $I$ into the 
parts $B$ and $A.$ We now see how to generalize the construction to make infinitely many examples of projectors $P$ such that $PP=P.$ To make such 
an example we choose a deformation $M$ of the identity line $I,$ and then cut it to obtain a factorization $I = BA.$ We then define $P$ by the 
equation $P = AB.$ See Figure~\ref{Figure 2.4} for an illustration of this process. There are infinitely many deformations of the identity, each giving rise to a
factorization via cutting the deformation in half, and each giving rise to distinct elements $P$ in the multiplicative Temperley Lieb algebra that 
have the property $PP=P.$ 
\bigbreak

We should point out that 
it is possible to generalize the element $I$ to $I_k$, a collection of $k$ parallel lines. Cutting a deformation of $I_k$ to obtain an factorization
$I_k = BA$ and an element $P = AB$ gives the general solution to the problem of finding all multiplicative elements in the Temperley Lieb algebra
with $PP=P.$ A modification of this approach yields a characterization of all elements $Q$ with $QQ = \delta^{r} Q$ for some positive integer $r.$
The basic structure behind this classification is the {\em meander}, a simple closed curve in the plane that has been bisected by a straight line.
Here we have illustrated the concept with an {\em open meander} consisting in a cutting of a deformation of the straight line $I.$ It is remarkable that
the classification of meanders is clearly formally related to the classification of folded molecules and, from this point of view, also related to 
the structure of self-reproduction. This section has been an abstract foray into the possibilities of topological genetics.
\bigbreak

\subsection{Protein Folding and Combinatorial Algebra}
\begin{figure}
     \begin{center}
     \begin{tabular}{c}
     \includegraphics[width=6cm]{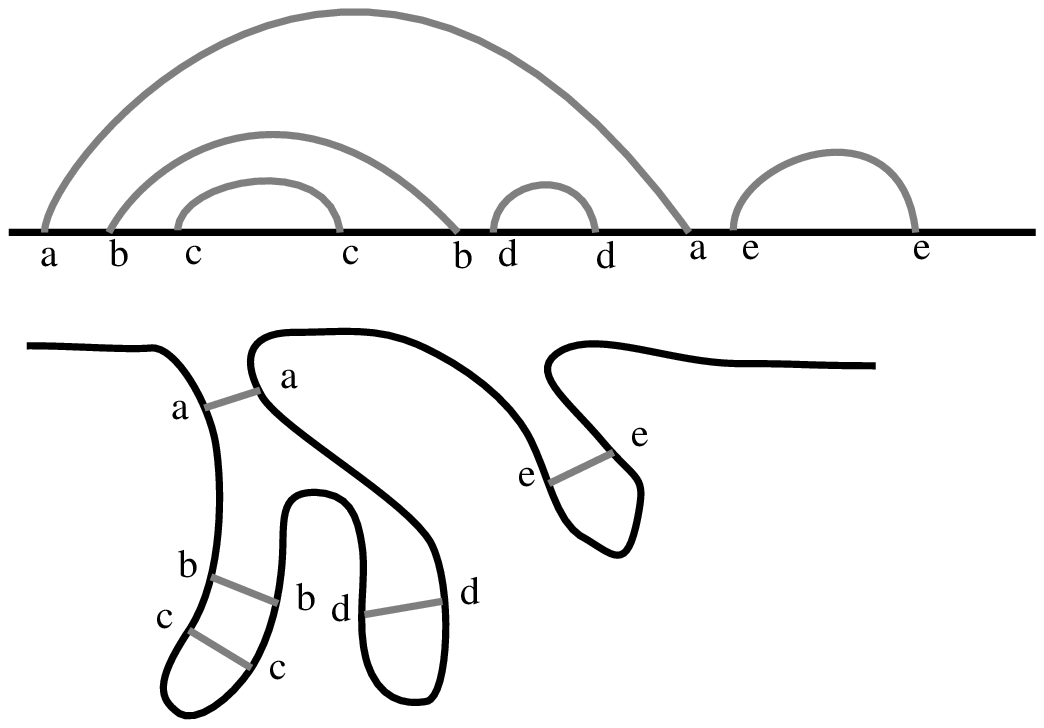}
     \end{tabular}
     \caption{\bf Secondary Structure $<a|<b|<c||c>|b><d||d>|a><e||e>$}
     \label{Figure 3}
\end{center}
\end{figure}

The approach in this section derives from ideas in \cite{KM}.
Here is another use for the formalism of bras and kets. Consider a molecule that is obtained by ``folding" a long chain molecule.
There is a set of sites on the long chain that are paired to one another to form the folded molecule. The difficult problem in protein 
folding is the determination of the exact form of the folding given a multiplicity of possible paired sites. Here we assume that the 
pairings are given beforehand, and consider the abstract structure of the folding {\em and} its possible embeddings in three dimensional
space. {\em Let the paired sites on the long chain be designated by labeled bras and kets with the bra appearing before the ket in the 
chain order.}  Thus $<A|$  and $|A>$ would denote such a pair and 
the sequence  $$C = <a|<b|<c||c>|b><d||d>|a><e||e>$$ \noindent could denote the paired sites on the long chain. See Figure~\ref{Figure 3} for 
a depiction of this chain and its folding.
In this formalism we do not assume any identities about moving containers or extainers, since the 
exact order of the sites along the chain is of great importance. We say that two chains are {\em isomorphic} if they differ only in
their choice of letters. Thus $<a|<b||b>|a>$ and $<r|<s||s>|r>$ are isomorphic chains. Note that each bra ket pair in a chain is decorated
with a distinct letter.
\bigbreak

Written in bras and kets a chain has an underlying parenthesis structure
that is obtained by removing all vertical bars and all letters. Call this $P(C)$ for a given chain $C$. Thus we have

$$P(C) = P(<a|<b|<c||c>|b><d||d>|a><e||e>) = <<<>><>><>.$$

Note that in this case we have $P(Chain)$ is a legal parenthesis structure in the usual sense of containment and paired brackets.
Legality of parentheses is defined inductively:
\begin{enumerate}
\item $<>$ is legal.
\item If $X$ and $Y$ are legal, then $XY$ is legal.
\item If $X$ is legal, then $<X>$ is legal.
\end{enumerate}
\noindent These rules define legality of finite parenthetic expresssions.
In any legal parenthesis structure, one can deduce directly from that structure which brackets are paired with one another. 
Simple algorithms suffice for this, but we omit the details. In any case a legal parenthesis structure has an intrinsic pairing associated with it,
and hence there is an inverse to the mapping $P$. We define $Q(X)$ for $X$ a legal parenthesis structure, to be the result of
replacing each pair $\cdots < \cdots > \cdots$ in X by $\cdots <A| \cdots |A> \cdots$ where $A$ denotes a specific letter chosen for that 
pair, with different pairs receiving different letters. Thus $Q(<<>>) = <a|<b||b>|a>.$ Note that in the case above, we have that 
$Q(P(C))$ is isomorphic to $C.$
\bigbreak

A chain $C$ is said to be a {\em secondary folding structure} if $P(C)$ is legal and $Q(P(C))$ is isomorphic to $C.$  
The reader may enjoy the exercise of seeing that 
secondary foldings (when folded) form tree-like structures without any loops or knots. This notion of secondary folding structure
corresponds to the usage in molecular biology, and it is a nice application of the bra ket formalism. This also shows the very rich
combinatorial background in the bras and kets that occurs before the imposition of any combinatorial algebra.
\bigbreak

\begin{figure}
     \begin{center}
     \begin{tabular}{c}
     \includegraphics[width=6cm]{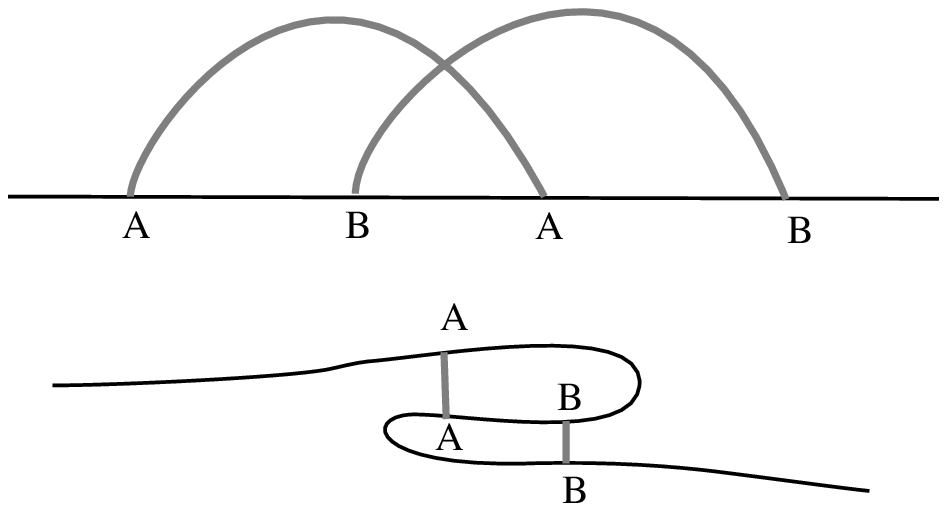}
     \end{tabular}
     \caption{\bf A Tertiary Structure - $<a|<b||a>|b>$}
     \label{Figure 4}
\end{center}
\end{figure}

Here is the simplest non-secondary folding: $$L = <a|<b||a>|b>.$$
\noindent Note that $P(L) =<<>>$ is legal, but that $Q(P(L)) = Q(<<>>) =<a|<b||b>|a>$ is not isomorphic to $L.$  
$L$ is sometimes called a ``pseudo knot" in the literature of protein folding. Figure~\ref{Figure 4} should make clear this nomenclature.
The molecule is folded back on itself in a way that looks a bit knotted.
\bigbreak

With these conventions it is convenient to abbreviate a chain by just giving its letter sequence and removing the (reconstructible) bras and kets.
Thus $C$ above may be abbreviated by $abccbddaee.$
\bigbreak

One may wonder whether at least theoretically there are foldings that would necessarily be knotted when embedded in three dimensional space.
With open ends, this means that the structure folds into a graph such that there is a knotted arc in the graph for some traverse from one end to 
the other. Such a traverse can go along the chain or skip across the bonds joining the paired sites.
The answer to this question is yes, there are folding patterns that can force knottedness. Here is an example of such an 
intrinsically knotted folding.

$$ABCDEFAGHIJKBGLMNOCHLPQRDIMPSTEJNQSUFKORTU.$$

It is easy to see that this string is not a secondary structure. To see that it is intrinsically knotted, we appeal to the Conway-Gordon Theorem
\cite{CG} that tells us that the complete graph on seven vertices is intrinsically knotted. In closed circular form (tie the ends of
the folded string together), the folding that 
corresponds to the above string retracts to the complete graph on seven vertices. Consequently, that folding, however it is embedded, must contain a knot
by the Conway-Gordon Theorem. We leave it as an exercise for the reader to draw an embedding corresponding to a folding of 
this string and to locate the knot!
The question of intrinsically knotted foldings that occur in nature remains to be investigated.
\bigbreak

\section{Recursive Distinguishing and Cellular Automata}
{\it Recursive Distinguishing (RD)} is a name coined by Joel Isaacson in his pioneering patent document \cite{JI} describing how fundamental patterns of process arise from the systematic 
application of operations of distinction and description upon themselves. \\

\noindent {\bf RD = [Distinction/Description Performed Recursively].}\\

In this section we will describe one version of Isaacson's RD process, and we will show how it gives rise to a pattern self-replication that is recognizable as a case of the replication that we have examined in this paper and called {\it DNA Replication}.  To find a process model of great simplicity that contains patterns that occur in complex systems is of great value to both the theoretican and the practical scientist. Such things lead to new insights, new questions and eventually to new applications.\\

The rules for the RD process are very simple. We begin with an arbitrary finite text string delimited by the character * at both ends. The RD process creates a new string from the given string by {\it describing the distinctions in the initial string.} Each character in the initial string is examined together with its
left and right neighbors. Let $LCR$ denote a character $C$ with neighbors $L$ and $R.$ Then we replace $C$ by a new character according to the following rules:
\begin{enumerate}
\item $C \longrightarrow  \, = \,$ if $L = C$ and $C = R$  (no distinction).
\item $C \longrightarrow   \, [ \,$ if $L \neq C$ but $C = R$ (distinction on the left).
\item $C \longrightarrow  \, ] \,$ if  $L = C$ but $C \neq R$ (distinction on the right).
\item $C \longrightarrow  \, o \,$ if  $L \neq C$ and $C \neq R$ (distinction on both the left and the right).
\item If $C$ is adjacent to $*$ change $C$ to $=$. (This is just a choice of boundary behavior.)
\end{enumerate}

\begin{figure}
     \begin{center}
     \begin{tabular}{c}
     \includegraphics[width=6cm]{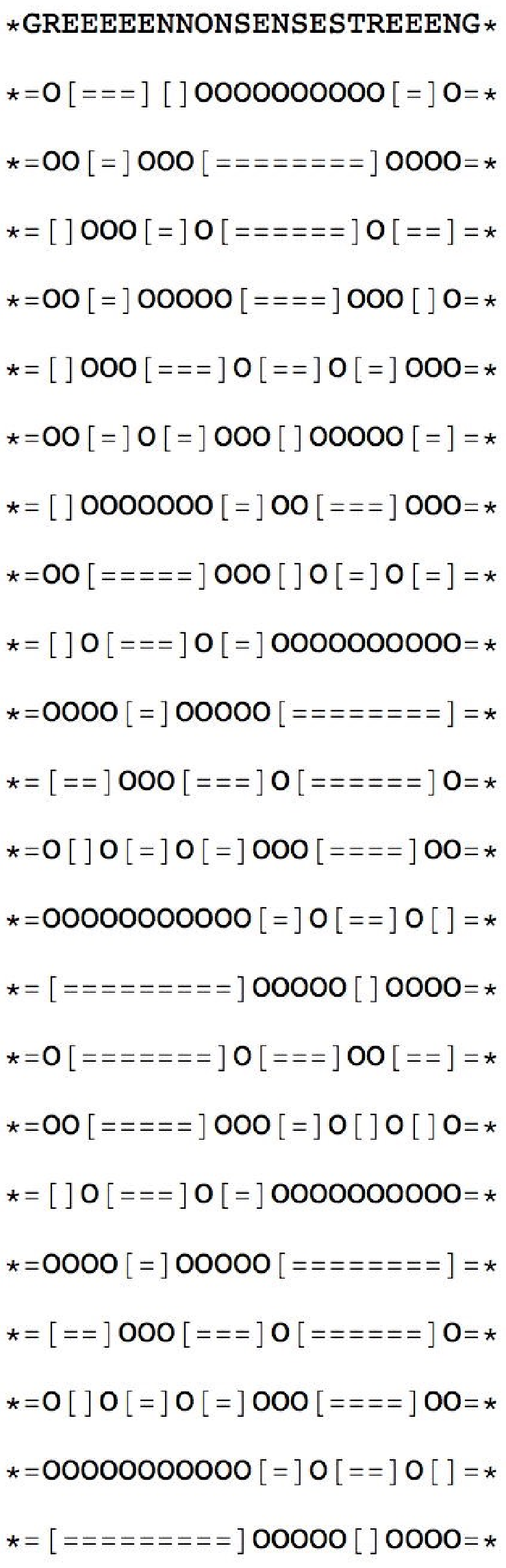}
     \end{tabular}
     \caption{\bf A String Evolution}
     \label{green}
\end{center}
\end{figure}

See Figure~\ref{green} for the result of applying the $RD$ process to a chosen text string.\\

\begin{figure}
     \begin{center}
     \begin{tabular}{c}
     \includegraphics[width=7cm]{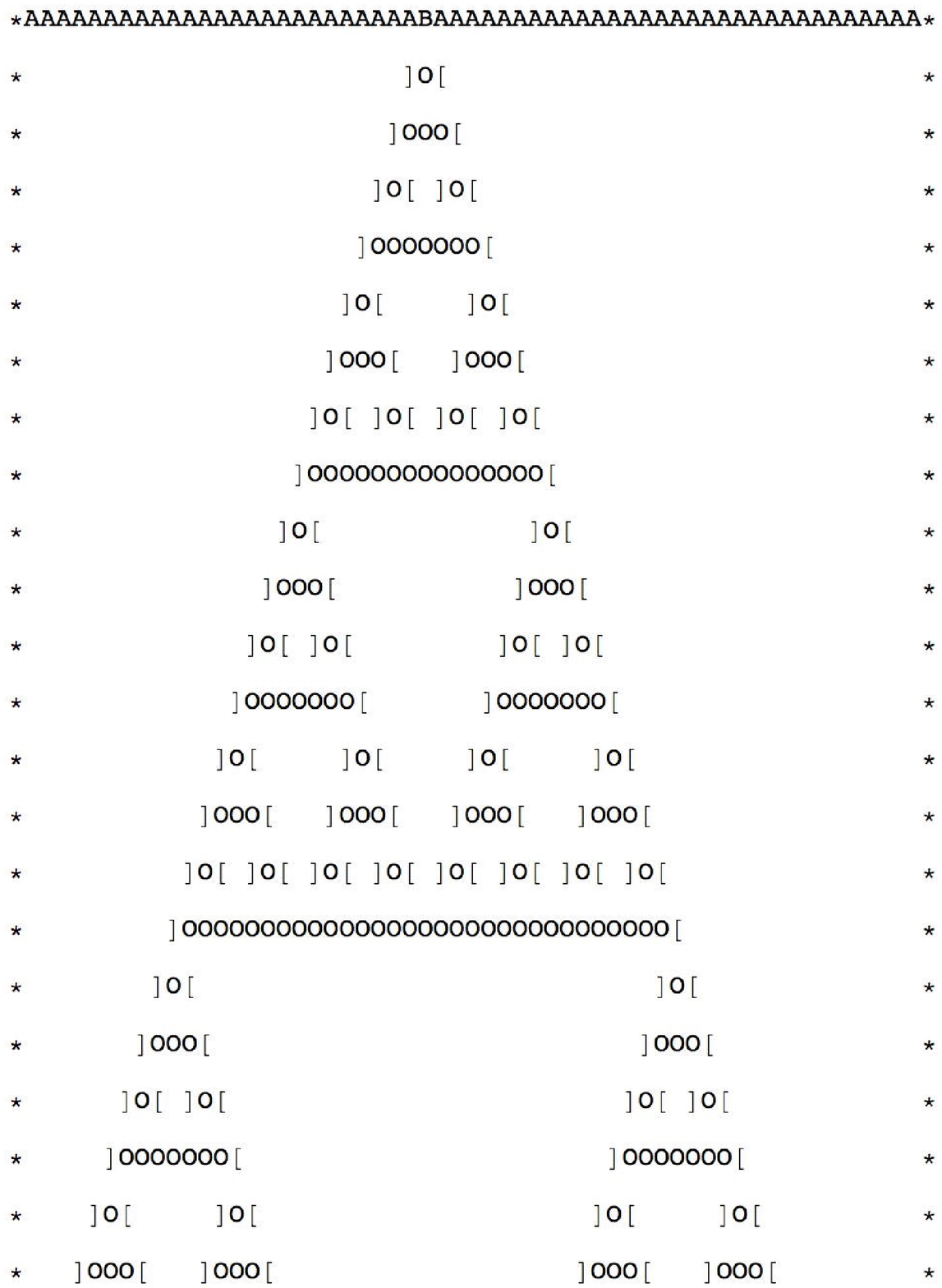}
     \end{tabular}
     \caption{\bf RD Replication}
     \label{rep}
\end{center}
\end{figure}

In Figure~\ref{rep} we show the result of starting with a very simple text string. In this figure we do not print the character $=$, so that the resulting strings have empty space where this character would appear. As the reader  can see, the string $*======]O[======*$
has a long sequence of transformations under the RD process. 
the pattern $]O[$ is replicated by the sequence below.\\
\begin{enumerate}
\item  $É=======]O[=======É.$
\item  $É======]OOO[=======É$
\item $ É=====]O[=]O[======É$
\end{enumerate}

Remarkably, this has the same patten as our previous description of abstract DNA replication.\\

We can interpret this as
\begin{enumerate}
\item $ ] = $ Watson, $[ \, =$ Crick, $O = $ backbone or binding.
\item  RD action results in the opening of the backbone so that binding $O$ is replaced by environment $OOO$.
\item  RD action relative to the environment results in placement of new Watson and new Crick. 
So we have the self-replication of $]O[$.
\end{enumerate}

In our previous story, we say  $DNA = WC = > <.$
This goes to $> E <$ where $E$ is the environment.
Then  $>E<$ goes to $ > < > < = DNA DNA.$
This is the general form of self-replication that applies to $DNA.$
Remarkably, the RD does exactly this in the formal sense. \\

Note that there is another level at which we can think about this! Regard $]$ and $[$ as ``cell-wallsÓ. Then we are witnessing not $DNA$ reproduction, but $MITOSIS$ itself!
The little fellow $]O[$ is a cell and we are watching how he reproduces in the line environment $ É=============É$. of the ``void" where there are no distinctions.
The reader should now look again at Figure~\ref{green} and note the many appearances and interactions related to this elementary cell.\\

Of course the interpretations of ``backbone", ``strand", ``environment" , ``cell"  are different from what happens in the biology, but it is very interesting that the basic principles are similar.\\

Note how we get  $ \cdots ===]OOOOO \cdots$ goes to   $ \cdots ==]O[=== \cdots$
So actually the whole ``environment" flips here.
But it is contained in the above scenario.
Everything that happens in RD is non-local since a single event affects the whole string.\\

Perhaps it is clear to the reader that Recursive Distinguishing in the sense of this section is a potentially explosive topic that will grow to influence all the aspects of biology and computing.
We believe that this is the case and we are in the process of writing other papers about it, including \cite{IJK}. The principle of [distinction/description in recursive process] applies at all levels of biology, coginition, information science and computing. \\

\subsection{Maturana, Uribe and Varela and the Game of Life}
\begin{figure}
     \begin{center}
     \begin{tabular}{c}
     \includegraphics[width=6cm]{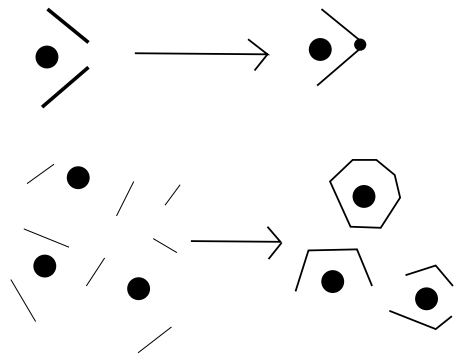}
     \end{tabular}
     \caption{\bf Proto-Cells of Maturana, Uribe and Varela}
     \label{Figure 5}
\end{center}
\end{figure}

Some examples from cellular automata clarify many of the issues about replication and the relationship of logic and biology. 
Here is an example due to 
Maturana, Uribe and Varela \cite{MUV}. See also \cite{FV} for a global treatment of related issues.
The ambient space is two dimensional and in it there are
``molecules" consisting in ``segments" and ``disks" (the catalysts) (See Figure~\ref{Figure 5}). There is a minimum distance among the 
segments and the disks (one can place them on a discrete lattice in the plane). And ``bonds" can form with
a probability of creation and a probability of decay between segment molecules with minimal
spacing. There are two types of molecules: ``substrate" (the segments) and ``catalysts" (the disks). The catalysts are not
susceptible to bonding, but their presence (within say three minimal step lengths) enhances the 
probability of bonding and decreases the probability of decay. Molecules that are not bonded
move about the lattice (one lattice link at a time) with a probability of motion.
In the beginning there is a randomly placed soup of molecules with a high percentage of substrate 
and a smaller percentage of catalysts. What will happen in the course of time?
\bigbreak

In the course of time the catalysts (basically separate from one another due to lack of bonding) 
become surrounded by circular forms of bonded or partially bonded substrate. A distinction 
(in the eyes of the observer) between inside (near the catalyst) and outside 
(far from a given catalyst) has spontaneously arisen through the ``chemical rules". Each catalyst
has become surrounded by a proto-cell. No higher organism has formed here, but there is a hint of the 
possibility of higher levels of organization arising from a simple set of rules of interaction.
{\em The system is not programmed to make the proto-cells.} They arise spontaneously in the evolution of 
the structure over time.
\bigbreak

One might imagine that in this way, organisms could be induced to arise as the evolutionary 
behavior of formal systems. There are difficulties, not the least of which is that there are 
nearly always structures in such systems 
whose probability of spontaneous emergence is vanishingly small. A good example is given by another 
automaton --  John H. Conway's ``Game of Life".  In ``Life" the cells  appear and disappear as
marked squares in a rectangular planar grid. A newly marked cell is said to be ``born". An unmarked cell
is ``dead". A cell dies when it goes from the marked to the unmarked state. A marked cell 
survives if it does not become unmarked in a given time step.
According to the rules of Life, an unmarked cell is born if and only if it has three neighbors.
A marked cell survives if it has either two or three neighbors. All cells in the lattice are updated in 
a single time step. The Life automaton is one of many automata of this type and indeed it is 
a fascinating exercise to vary the rules and watch a panoply of different behaviors. 
\bigbreak

\begin{figure}
     \begin{center}
     \begin{tabular}{c}
     \includegraphics[width=6cm]{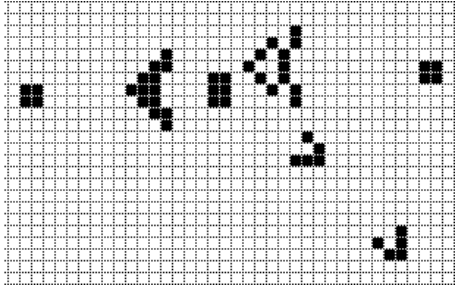}
     \end{tabular}
     \caption{\bf Glider Gun and Gliders}
     \label{Figure 6}
\end{center}
\end{figure}

For this 
discussion we concentrate on some particular features. There is a configuration in Life called a
``glider". See Figure~\ref{Figure 6}, which illustrates a series of gliders
going diagonally from left to right down the Life lattice, as well as a "glider
gun" (discussed below) that has produced them.
The glider consists in five cells in one of two basic configurations.
Each of these configurations produces the other (with a change in orientation). After four steps the
glider reproduces itself in form, but shifted in space. Gliders appear as moving entities in 
the temporality of the Life board. The glider is a complex entity that arises naturally from a 
small random selection of marked cells on the Life board. Thus the glider is a ``naturally 
occurring entity" just like the proto-cell in the Maturana-Uribe-Varela automaton. 
\bigbreak

But Life 
contains potentially much more complex phenomena. For example, there is the ``glider gun" (See
Figure~\ref{Figure 6}) which perpetually creates new gliders.  The ``gun" was invented by the Gosper Group, a group of researchers
at MIT in the 1970's. It is highly unlikely that a gun would appear 
spontaneously in the Life board. Of course there is a tiny probability of this, but we would guess
that the chances of the appearance of the glider gun by random selection or evolution from a 
random state is similar to the probability of all the air in the room collecting in one corner.
Nervertheless, the gun is a natural design based on forms and patterns that do appear spontaneously 
on small Life boards.  The glider gun emerged through the coupling of the power of human cognition and the automatic behavior of 
a mechanized formal system.
\bigbreak
  
Cognition is in fact an attribute of our biological system at an 
appropriately high level of organization. Cognition itself looks as improbable as the glider gun!  
Do patterns as complex as cognition or the glider gun arise 
spontaneously in an appropriate biological context? 
\bigbreak

There is a middle ground.  If one examines cellular automata of a given type and varies the
rule set randomly rather than varying the initial conditions for a given automaton, then a very wide
variety of phenomena will present themselves. In the case of molecular biology at the level of the 
DNA there is exactly this possibility of varying the rules, in the sense of varying the sequences in 
the genetic code. So it is possible at this level to produce a wide range of remarkable complex systems.
\bigbreak

\subsection{Other Forms of Replication}

Other forms of self-replication are quite revealing. For example, one might point out that a
stick can be made to reproduce by breaking it into two pieces. This may seem satisfactory on the 
first break, but the breaking cannot be continued indefinitely. In mathematics on the other hand, 
we can divide an interval into two intervals and continue this process ad infinitum. For a 
self-replication to have meaning in the physical or biological realm there must be a genuine 
repetition of structure from original to copy. At the very least the interval should grow to 
twice its size before it divides (or the parts should have the capacity to grow independently).
\bigbreak

\begin{figure}
     \begin{center}
     \begin{tabular}{c}
     \includegraphics[width=6cm]{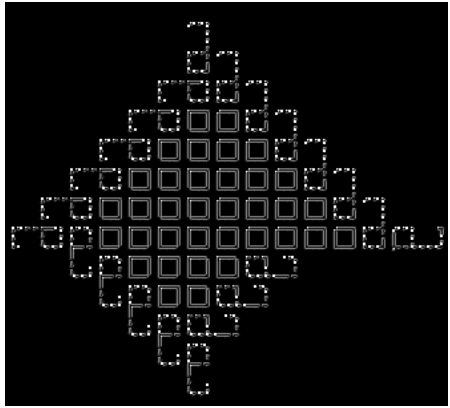}
     \end{tabular}
     \caption{\bf Langton's Automaton}
     \label{Figure 7}
\end{center}
\end{figure}

A clever automaton, due to Chris Langton, takes the initial form of a square in the plane.
The square extrudes an edge that grows to one edge length and a little more, 
turns by ninety degrees, grows one edge length, turns by ninety degrees grows one edge length,
turns by ninety degrees and when it grows enough to collide with the original extruded edge, 
cuts itself off to form a new adjacent square, thereby reproducing itself. This scenario is 
repeated as often as possible producing a growing cellular lattice. See Figure~\ref{Figure 7}. 
\bigbreak

\begin{figure}
     \begin{center}
     \begin{tabular}{c}
     \includegraphics[width=6cm]{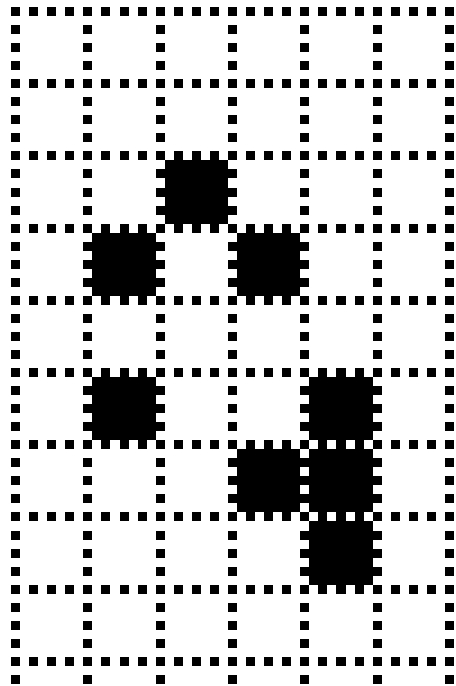}
     \end{tabular}
     \caption{\bf Condition D with geometric period $48$}
     \label{Figure 8}
\end{center}
\end{figure}

The replications that happen in automata such as Conway's Life are all really instances of periodicity
of a function under iteration. The glider is an example where the Life game function $L$ applied to an
initial condition $G$ yields $L^{5}(G) = t(G)$ where $t$ is a rigid motion of the plane. Other intriguing
examples of this phenomenon occur. For example the initial condition $D$ for Life shown in Figure~\ref{Figure 8} has the property that 
$L^{48}(D) = s(D) + B$ where $s$ is a rigid motion of the plane and $s(D)$ and the residue $B$ are disjoint
sets of marked squares in the lattice of the game. $D$ itself is a small configuration of eight marked
squares fitting into a rectangle of size $4$ by $6.$ Thus $D$ has a probability of $1/735471$ of being chosen
at random as eight points from $24$ points.  
\bigbreak

\begin{figure}
     \begin{center}
     \begin{tabular}{c}
     \includegraphics[width=6cm]{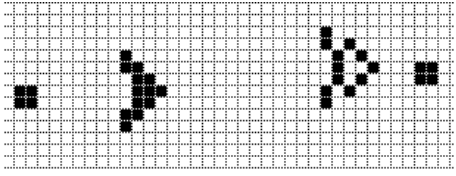}
     \end{tabular}
     \caption{\bf P(left) and Q(right) Compose the Glider Gun}
     \label{Figure 9}
\end{center}
\end{figure}

Should we regard self-replication as simply an instance of periodicity under iteration?
Perhaps, but the details are more interesting in a direct view.  The glider gun in Life is a
structure $GUN$ such that $L^{30}(GUN) = GUN + GLIDER.$ Further iterations move the disjoint glider away
from the gun so that it can continue to operate as an initial condition for $L$ in the same way.
A closer look shows that the glider gun is fundamentally composed of two parts $P$ and $Q$ such that 
$L^{10}(Q)$ is a version of $P$ and some residue and such that $L^{15}(P) = P^* + B$ where $B$ is a rectangular 
block, and $P^*$ is a mirror image of $P$, while $L^{15}(Q) = Q^* + B'$ where $B'$ is a small non-rectangular
residue. See Figure~\ref{Figure 9} for an illustration showing the parts $P$ and $Q$ (left and right) flanked by small blocks that
form the ends of the gun. One also finds that $L^{15}( B + Q^*) = GLIDER + Q + Residue.$ This is the internal 
mechanism by which the glider gun produces the glider. The extra blocks at either end of the glider
gun act to absorb the residues that are produced by the iterations. Thus the end blocks are catalysts
that promote the action of the gun. Schematically the glider production goes as follows:

$$P+Q \longrightarrow P^* + B + Q^*$$

$$B + Q^* \longrightarrow GLIDER + Q$$

\noindent whence

$$P+Q \longrightarrow P^* + B + Q^* \longrightarrow P + GLIDER + Q = P + Q + GLIDER.$$

\noindent The last equality symbolizes the fact that the glider is an autonomous entity no longer involved 
in the structure of $P$ and $Q.$ It is interesting that $Q$ is a spatially and time shifted version of 
$P.$ Thus $P$ and $Q$ are really ``copies" of each other in an analogy to the structural relationship of 
the Watson and Crick strands of the DNA. The remaining part of the analogy is the way the catalytic
rectangles at the ends of the glider gun act to keep the residue productions from interfering with 
the production process. This is analogous to the enzyme action of the topoisomerase in the DNA.
\bigbreak

The point about this symbolic or symbiological analysis is that it enables us to take an analytical 
look at the structure of different replication scenarios for comparison and for insight.
\bigbreak

\section{Epilogue - Logic, Biology and Phenomenology}

We began with the general question: What is the relationship of logic and biology and how can these apparently separate points of view be embraced in phenomenology.
Certain fundamentals, common to both, are handled quite differently. There are certain fundamental distinctions
between  symbol and object (the name and the thing that is named), and between a
form and a copy of that form. 
\bigbreak

In logic the symbol and its referent are normally taken to be distinct. Nevertheless we have seen that at the basis of such systems symbols may be self-referent, just as
a bracket $\langle \rangle$ refers to enclosure and is itself an enclosure. This leads to a host of 
related distinctions such as the distinction between a description or blueprint and the object 
described by that blueprint.  A related distinction is the dichotomy between software and hardware.
The software is analogous to a description. Hardware can be constructed with the aid of a blueprint or
description. But software intermediates between these domains as it is an {\em instruction.} An instruction is not
a description of a thing, but a blueprint for a process. Software needs hardware in order to 
become an actual process. Hardware needs software as a directive force. Although mutually dependent,
hardware and software are quite distinct.
\bigbreak

In logic and computer science the boundary between hardware and software is first met at the machine 
level with the built-in capabilities of the hardware determining the type of software that can be 
written for it. Even at the level of an individual gate, there is the contrast of the 
structure of that gate as a design and the implementation of that design that is used in the
construction of the gate. The structure of the gate is mathematical. Yet there is the physical implementation of these
designs, a realm where the decomposition into parts is not easily 
mutable. Natural substances are used, wood, metal, particular compounds, atomic elements and so on.
These are subject to chemical or even nuclear analysis and production, but eventually one 
reaches a place where Nature takes over the task of design.
\bigbreak

In biology it is the reverse. No human hand has created these designs. The organism stands for itself,
and even at the molecular level the codons of the DNA are not symbols. They do not stand for 
something other than themselves. They cooperate in a process of production, but no one wrote their sequence as
software. There is no software. There is no distinction between hardware and software in 
biology. 
\bigbreak

Nevertheless, one cannot resist pointing out that with DNA each base, potentially paired as it is with its complementary base, can be regarded as a {\it character in
a text string} that is ``described'' by its complementary base. The base pairs are $AT$ (Adenine and Thymine) and $GC$ (Guanine and Cytosine). We can regard the rule that replaces
$A$ by $T$ , $T$ by $A$, $G$ by $C$ and $C$ by $G$ as the fundamental recursive distinction (RD)  (See Section 10) for the Watson and Crick strands. Each strand is the RD production of the other.
This mutual reference of the Watson and Crick strands is the fundamental entwining of the DNA , and it is accompanied by a topological entwinement as well. This double entwinement leads into all the complexities of biology. We see that every theme in this paper, of mathematics, topology, form, recursion and logic is reflected in the structure and process of the DNA.\\ 

\subsection{Syntax and Semantics}
 
In logic a form arises via the syntax and alphabet of a given formal system.  
That formal system arises via the choices of the mathematicians who create it. They create it through
appropriate abstractions. Human understanding fuels the operation of a formal system.
Understanding imaged into programming fuels the machine operation of a mechanical image of that formal 
system. The fact that both humans and machines can operate a given formal system has lead to much confusion, for they
operate it quite differently. 
\smallbreak

{\em Humans are always on the edge of breaking the rules either through error or inspiration.
Machines are designed by humans to follow the rules, and are repaired when they do not do so. Humans are encouraged to operate 
through understanding, and to create new formal systems (in the best of all possible worlds).} 
\smallbreak

Here is the ancient 
polarity of syntax (for the machine) and semantics (for the person). The person must mix syntax and semantics to come to
understanding. Here is the locus for deeper discussion of the role of phenomenology in logic and in the relationship of persons to logic. For we do not adhere to rules from the outside. 
Rather, we move to become actors for whom those rules are as natural as our own biology. Then the rules emanate from our doing and not so much followed as acknowledged.
\bigbreak

The movement back and forth between syntax and semantics underlies all attempts to
create logical or mathematical form. This is the cognition behind a given formal 
system. There are those who would like to create cognition on the basis of syntax alone.
But the cognition that we know is a byproduct or an accompaniment to biology. 
This is not to assert that biology is prior to cognition, but rather that the full absorbtion of following rules is a condition where the rules themselves come from our own actions in the world.
Biological cognition comes from a domain where there is at base no distinction between syntax 
and semantics. To say that there is no distinction between syntax and semantics in biology
is not to say that it is pure syntax. Syntax is born of the possibility of such a 
distinction.
\bigbreak

In biology an energetic chemical and quantum substrate gives rise to a ``syntax" of combinational 
forms (DNA, RNA, the proteins, the cell itself, the organization of cells into the organism).
These combinational forms give rise to cognition in human organisms. Cognition gives rise to
the distinction of syntax and semantics. Cognition gives rise to the possibility of design,
measurement, communication, language, physics and technology.
\bigbreak

\subsection{Extainers, Containers and Goedelian Reference}
In the discussion with Cookie and Parabel in Section 6. We introduced formalism leading to self-reference in the form of a Goedelian shift and the identification of 
the linguistic I with ``the meta-name of the meta-naming operator". This is both a strict bit of formalism showing how a system that handles reference and can refer to its own operations will naturally create a symbol for itself that is ceremonially both an indicator of name and process of naming. In this nexus of syntax and semantics there is the possibility for further progress in the understanding of 
how organism and awareness are intertwined.\\

In this paper we have covered a wide ground of ideas related to the foundations of mathematics and its relationship with biology and with 
physics. There is much more to explore in these domains. The result of our exploration has been the articulation of a mathematical region that 
lies in the crack between set theory and its notational foundations. We have articulated the concepts of container $<>$ and extainer $><$
and shown how the formal algebras generated by these forms encompass significant parts of the logic of DNA replication, the Dirac formalism for
quantum mechanics, formalism for protein folding and the Temperley Lieb algebra at the foundations of topological invariants
of knots and links.  It is the 
mathematician's duty to point out formal domains that apply to a multiplicity of contexts. In this case we suggest that it is just possible that 
there are deeper connections among these apparently diverse contexts that are only hinted at in the steps taken so far. The common formalism 
can act as compass and guide for further exploration.
\bigbreak 

I would like to end with an example that is purely diagrammatic.  It is well-known that Goedel's incompleteness theorem is based on a method of coding expressions in a formal system by natural numbers in such a way that one can create statements that refer to themselves by referring to their own code numbers. I wanted to diagram the Goedelian situation to clarify it for myself.
Accordingly, I chose \cite{Arrow} an arrow of reference in the most general sense.
$$A \longrightarrow B$$

The arrow will mean that ``A refers to B".  One way to refer is to name, and so the arrow can be interpreted to mean that ``A is the name of B." Thus if $g$ is a Goedel number, then 
$$g \longrightarrow F(u)$$
can mean that ``g is the Goedel number of the formula F(u)".  Now with the help of the arrow, we can diagram the famous shift that Goedel devised. Let us assume that the formula $F(u)$ has a single free variable $u.$ (It might be something like ``$u$ is a prime number".) Goedel invented a function that I shall denote by $ \sharp g.$  This function is described by the statement ``$\sharp g$ is the Goedel number of the formula obtained from $F(u)$ by substituting the Goedel number of $F(u)$ into the free variable of $F(u).$". That mouthful becomes the following arrow diagram:
$$g \longrightarrow F(u),$$ then
$$\sharp g \longrightarrow F(g).$$

\noindent Let us call the second arrow the Goedelian shift of the first arrow.
Now comes the amazing construction of Goedel. Suppose we have a formula that uses the function $\sharp u.$  It can have the form$ F(\sharp u)$ and it has Goedel number equal to $ g.$
So we have 
$$g \longrightarrow F(\sharp u)$$  
shifting to
$$\sharp g \longrightarrow F(\sharp g).$$

Mirablile dictu, the formula $F(\sharp g)$ is discussing its own Goedel number!
This is the key to the Goedelian construction of self-reference. It is the heart of the incompletenesss theorem whereby Goedel creates a sentence that asserts its own unprovability in the given formal system. \\

The arrow for reference makes the logic of this profound construction easy to survey, and it makes it possible to put the construction in a more general context to see that Goedelian self-reference is very like what we do in language with naming, and eventually naming ourselves. It is this general context that Cookie and Parbel discuss near the end of their conversation in our section on the 
``Epistemology of Text-Strings". Cookie and Parabel discuss a shift of a more general kind where
$$ A \longrightarrow B$$ shifts to $$\sharp A \longrightarrow BA.$$ Here we regard the arrow as an arrow of reference as before and the shift appends the ``name" A of B to B to form a new text-string
AB. Just as in Goedel, self reference occurs when we shift the name of the meta-naming operator.
$$M \longrightarrow \sharp$$ shifts to 
$$M \sharp \longrightarrow M \sharp.$$ 
This is the form of self-reference arising in a putative organism that can name and refer and shift the reference. The shifting of reference is something we do all the time at the linguistic level. We place the name of the object with the object in the cognitive space. And when we place the name of our own operation of naming with that operation we construct the abstract precursor to an I. We identify our name with the process that we are. Just as with replication of DNA this is a schema for the self. We have ignored everything that was important except for the bare-bones of the process of reference. In so doing we can draw a parallel between the syntactical patterns of self-reproduction at the molecular level and the syntactical patterns of self-reference, usually imagined at the level of mind and language. It is no coincidence that we find them 
together here, since we have allowed it. The speculation of this paper is that indeed there is no real distance between the nuts and bolts of molecular biology and the semantics of mind.\\ 

\noindent {\bf Acknowledgment.} The author thanks Sofia Lambropoulou for many useful conversations
in the course of preparing this paper. The author also thanks Sam Lomonaco, John Hearst, Yuri Magarshak, James Flagg, William Bricken and Joel Isaacson
for conversations related to the content of the present paper.
\smallbreak

\bigbreak

\end{document}